\documentclass{article} %
\usepackage{main,times}

\usepackage{bbm}
\usepackage{subcaption} 

\usepackage{amsmath,amsfonts,bm}

\def\eqref#1{equation~\ref{#1}}

\def\1{\bm{1}}

\DeclareMathAlphabet{\mathsfit}{\encodingdefault}{\sfdefault}{m}{sl}
\SetMathAlphabet{\mathsfit}{bold}{\encodingdefault}{\sfdefault}{bx}{n}

\usepackage{graphicx}
\usepackage{array}
\usepackage{amsmath}
\usepackage{multirow}
\usepackage{booktabs}
\usepackage{hyperref}
\usepackage{url}
\usepackage{wrapfig,lipsum,booktabs}
\usepackage{xspace}
\usepackage{caption}

\usepackage{placeins}
\usepackage[bottom]{footmisc}

\title{DELPHYNE: A Pre-Trained Model for General and Financial Time Series}

\author{Xueying Ding \\ Carnegie Mellon University \\ xding2@cs.cmu.edu \And Aakriti Mittal \\ Bloomberg  \\ amittal114@bloomberg.net \And Achintya Gopal \\ Bloomberg\thanks{Work done while employed at Bloomberg.} \\achintyagopal@gmail.com}

\newcommand{\methodl}{{\small\sf Delphyne-L}\xspace}
\newcommand{\methoda}{{\small\sf Delphyne-A}\xspace}
\newcommand{\methodf}{{\small\sf Delphyne-F}\xspace}

\iclrfinalcopy %
\begin{document}

\maketitle

\begin{abstract}
Time-series data is a vital modality within data science communities. This is particularly valuable in financial applications, where it helps in detecting patterns, understanding market behavior, and making informed decisions based on historical data. Recent advances in language modeling have led to the rise of time-series pre-trained models that are trained on vast collections of datasets and applied to diverse tasks across financial domains. However, across financial applications, existing time-series pre-trained models have not shown boosts in performance over simple finance benchmarks in both zero-shot and fine-tuning settings. This phenomenon occurs because of a i) lack of financial data within the pre-training stage, and ii) the negative transfer effect due to inherently different time-series patterns across domains. Furthermore, time-series data is continuous, noisy, and can be collected at varying frequencies and with varying lags across different variables, making this data more challenging to model than languages. To address the above problems, we introduce a Pre-trained Mo\textbf{DEL} for \textbf{FIN}ance Tim\textbf{E}-series (\textbf{Delphyne}). \textbf{Delphyne} achieves competitive performance to existing foundation and full-shot models with few fine-tuning steps on publicly available datasets, and also shows superior performances on various financial tasks.
\end{abstract}

\section{Introduction}
\label{sec:intro}
Time series is one of the most ubiquitous modalities in finance. Time-series analysis is critical to various tasks, such as asset pricing, volatility modeling, risk management, economic indicator analysis, etc. Traditional statistical methods for financial time series include Autoregressive Integrated Moving Average (ARIMA), Generalized Autoregressive Conditional Heteroskedasticity (GARCH), and Vector Autoregressive (VAR) models. In recent years, deep learning-based methods are being applied to these financial tasks (e.g., \cite{horvath2019fastpricing,Araujo2023InflationForecast,liu2024neuralbetaestimatingbetausing,gopal2024neuralfactorsnovelfactorlearning}). 

Following the surge of large language models (LLMs), research has explored the possibility of large deep-learning models for time-series analysis. One approach focuses on prompting and adapting existing pre-trained language models for time-series tasks \citep{zhou2023ofa,jin2024timellm,chang2024llm4ts,gruver2024LLMTime}, while another line of research  trains foundation models directly on time-series data \citep{woo2024unifiedtraininguniversaltime,das2024timesfm,goswami2024moment}. Both approaches aim to train a single model with large-scale cross-domain data that can be applied to various tasks, as opposed to full-shot methods that require training a distinct model for each dataset.

However, previous research indicates that directly prompting LLMs for financial tasks brings only modest benefits; moreover, most LLM-based research focuses only on forecasting directional trends of stock returns, offering fewer practical applications than traditional methods \citep{Chen2023chatgptstock,yu2023temporaldatameetsllm,lopezlira2023chatgptforecaststockprice}. Preliminary experiments also reveal that pre-trained models do not demonstrate significant improvement over GARCH in risk analysis (Sec. \ref{sec:exp}).

One reason for this limitation is that the existing public datasets for pre-training lack sufficient financial data to learn patterns and distributional properties that are specific to time-series data in finance. 
Jointly training on financial data alongside public datasets presents an additional challenge: the substantial differences between domains can lead to \textit{negative transfer}. Negative transfer has been extensively discussed within the literature, defined as a case when ``transferring knowledge from the training data has an negative impact on the target
tasks''  \citep{wang2019characterizingavoidingnegativetransfer}. Negative transfer is identified as a key obstacle towards pre-training graph foundation model \citep{wang2024subgraphpoolingtacklingnegative,mao2024positiongraphfoundationmodels}, yet is not explored within the context of time-series data. In Sec. \ref{sec:negative_transfer}, we show that the negative transfer effect is a real challenge when pre-training time-series models. Cross-domain transfer learning is particularly difficult, as time-series data tends to be noisier and more continuous compared to languages and images. To alleviate the negative transfer effect, we believe that fine-tuning is the only remedy. We contend that the strength of pre-trained time-series models lies in their capacity to rapidly ``unlearn'' the biases of the pre-training stage and ``adapt'' to the specific distribution of new tasks, given limited training data and time.

Building on our observation and analysis of negative transfer, we introduce our pre-trained time-series model, which incorporates several architectural modifications designed to better handle time-series data and tasks. Time-series data often involve an arbitrary number of variates, each collected at different frequencies (e.g., where quarterly profit growth is forecasted from daily or weekly sales data). Additionally, finance tasks often involve nowcasting, where some variates contain contemporaneous data.
Our model is an encoder-based transformer that employs the any-variate attention mechanism, introduced by \cite{woo2024unifiedtraininguniversaltime}. In addition, we implement a missing data mask alongside the forecast mask, which enhances the model's ability to adapt to cross-frequency training while minimizing negative interference \citep{vanness2023crossfrequencytimeseriesmetaforecasting}.

Probabilistic forecasting is an essential feature required for uncertainty quantification and anomaly detection within time series. Existing literature either assumes a fixed output distribution \citep{salinas2019deeparprobabilisticforecastingautoregressive} or utilizes a mixture of several distributions \citep{woo2024unifiedtraininguniversaltime}. However, when the underlying data exhibits contrasting distributions and supports, a single distribution may be inadequate. Our model simplifies the output by using a mixture of Student-T distributions, which are widely used in finance for heavy-tailed tasks. In our  preliminary studies, this approach shows comparable performance to \cite{woo2024unifiedtraininguniversaltime} on downstream tasks.

In Sec. \ref{sec:method}, we introduce additional architectural modifications and present corresponding ablation studies. To this end, we train our Pre-Trained Mo\textbf{DEL} for \textbf{FIN}ance Tim\textbf{E}-series (\textbf{Delphyne}), the first time-series model capable of both general and finance-specific tasks. Delphyne is trained on both LOTSA (the largest public times-series dataset \citep{woo2024unifiedtraininguniversaltime}) and financial data (Sec.\ref{sec:training}). We conduct experiments for both general and finance time-series tasks, demonstrating that Delphyne consistently outperforms or matches state-of-the-art baselines. Our key contributions are as follows:

\vspace{-0.05in}
\begin{itemize}
    \item Demonstrate the presence of negative transfer in pre-trained time-series models, highlighting how this effect sets them apart from LLMs, emphasizing that the value of pre-trained time-series models lies in their fine-tuning capability, which mitigates the negative transfer effect and enhances downstream performance with minimal iterations.
    \item Introduce several architectural modifications in our model and demonstrate their significance through comprehensive ablation studies.
    \item Delphyne is the first time-series pre-trained model to excel in both general time-series tasks and a variety of downstream finance tasks.
\end{itemize}

\section{Negative Transfer Effect}
\label{sec:negative_transfer}

Negative transfer has been widely studied in foundation models across different modalities \citep{wang2024subgraphpoolingtacklingnegative}. This problem is typically seen as the model's reduced performance on downstream tasks due to mismatches between the source training data and the target distribution \citep{wang2019characterizingavoidingnegativetransfer}. However, this phenomenon has not been thoroughly explored in the context of time series. In pre-trained time-series models, negative transfer can occur when cross-domain data is added during pre-training. The appearances of data from too different distributions can lead to less effective zero-shot forecast results, even if we feed the downstream tasks within similar domains. We present three examples to highlight the presence of negative transfer, particularly focusing on the difficulties of cross-domain transfer from finance data to other areas.

\begin{wrapfigure}{r}{0.38\textwidth}
    \begin{minipage}{0.38\textwidth}
        \centering
        \includegraphics[width=0.48\textwidth]{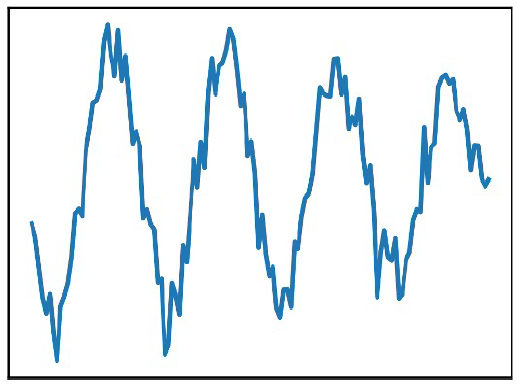}
        \includegraphics[width=0.48\textwidth]{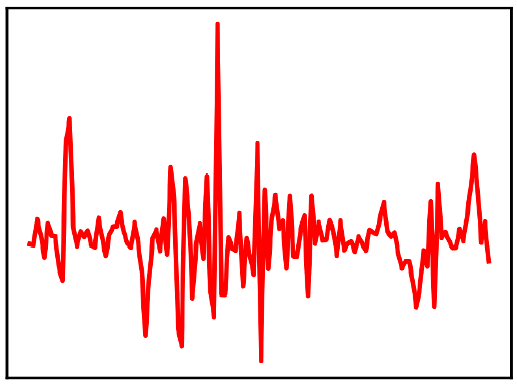}
        \vspace{-0.1in}
        \caption{(Left) Wavelet Function. (Right) GARCH-style data.}
        \label{fig:garch_vs_wavelet}
    \end{minipage}
    \hfill
    \begin{minipage}{0.38\textwidth}
        \centering
        \vspace{0.12in}
        \captionof{table}{Zero-shot NLL($\downarrow$) for models trained on different data types.}
        \vspace{-0.22in}
        \label{tab:negative-transfer-zeroshot}
        \scalebox{0.75}{
        \begin{tabular}{ccc}\\\toprule  
        Model & Wavelet Pred. & GARCH Pred. \\\midrule
        G-Model-128 & - & \textbf{0.0865} \\ 
        W-Model-128 & \textbf{-0.1020}  & - \\  
        M-Model-128 & -0.0733 & 0.1137 \\  \bottomrule
        G-Model-32 & - & \textbf{0.0882} \\ 
        W-Model-32 & \textbf{-0.1330}  & - \\  
        M-Model-32 & -0.0732 & 0.1176 \\  \bottomrule
        \end{tabular}}
    \end{minipage}
    \vspace{-0.2in}
\end{wrapfigure}

\textbf{Pre-training with GARCH and Wavelet Data} 
To simulate datasets encountered in real-world scenarios, we generate two types of synthetic data: Wavelet functions and GARCH-style data. Wavelet functions are composed of a combination of sine and cosine waves, while in GARCH models the current time-steps are based on past squared residuals and past variances, frequently used to capture  volatility clustering seen in financial time-series data. These data types are prevalent in nature and are standard simulations for time series in both finance and non-finance literature \citep{GARCH,das2024timesfm,Petrozziello2022GARCH}. Fig. \ref{fig:garch_vs_wavelet} shows examples of these two data types.

We trained three distinct models: the first on GARCH-style data (G-Model), the second on Wavelet data (W-Model), and the third on a combination of both, using half Wavelet and half GARCH data (M-Model). Each model utilizes a standard autoregressive transformer decoder without any additional embeddings or patching. Further details can be found in the Appendix \ref{ssec:pretrain_with_garch}. Table \ref{tab:negative-transfer-zeroshot} presents the negative log-likelihood (NLL) for each model's zero-shot forecasts with context lengths of $32$ and $128$. As expected, the mixed-data model (M-Model), trained on both GARCH and Wavelet data, performs significantly worse in terms of zero-shot NLL compared to the models trained on a single data source. This underperformance is likely due to the differing frequencies and structures of GARCH and Wavelet data.

\textbf{Bayesian MCMC}
While the previous example showcases that model trained on both GARCH and Wavelet data underperforms compared to models trained on a single type, one may argue that it is a model training problem. However, we provide further evidence that negative transfer occurs because of the contrasting properties of the data, regardless of the model used. 

We analyze the time-series pre-trained models in a Bayesian framework, in analogy to the Bayesian inference mechanism in \citep{muller2022transformerscandobayesianinference,hollmann2023tabpfntransformersolvessmall}. The pre-training stage is essential to fit a good prior to the training tasks. Given $N$ tasks (datasets) $\{\mathcal{D}_i\}_{i=1}^N$, the learned parameters $\mathbf{\hat{w}}$ implicitly presents the maximum likely distribution over all observed tasks by minimizing the pre-training objective. As shown by \citet{muller2022transformerscandobayesianinference}, optimizing the following training loss approximates the true Bayesian posterior forecast distribution for time series:
\begin{align}
\mathbb{E}_{ x_{1:t+1} \sim p(D)} \left[-\log q_{\theta} (x_{t+1} | x_{1:t}) \right]
\end{align}
In light of this framework, we can simulate the results of zero-shot learning by setting the prior to some data generating process (instead of learning from data). To illustrate our approach, we first construct a model to simulate the data-generating processes of Wavelet functions, and then generate corresponding observed data. 
We subsequently apply Markov Chain Monte Carlo (MCMC) to obtain posterior samples for the model (the distribution of model parameters after conditioning on the observed data).
Note that the Wavelet model is fully aware of the underlying data-generating process, and it should be capable of modeling the distribution accurately, apart from the additive Gaussian noise. Then, we fit a mixture of Wavelets and GARCH to the same data, see Appendix \ref{ssec:bayesian_mcmc}. 

\begin{wraptable}{r}{4.5cm}
\vspace{-0.15in}
\caption{Forecasted NLL($\downarrow$) and MSE($\downarrow$) for Wavelet-MCMC, and Mixture-MCMC model.}\label{tab:bayeisan_mcmc_result}
\vspace{-0.2in}
\scalebox{0.8}{
\begin{tabular}{ccc}\\\toprule  
Model & MSE. & NLL. \\\midrule
Wavelet-MCMC & \textbf{0.1058} & \textbf{0.2692} \\ 
Mixture-MCMC & 0.1127  & 0.2937 \\ \bottomrule
\end{tabular}
}
\end{wraptable}

We compute the NLL and Mean Squared Error (MSE) for the fitted Wavelet-MCMC model and the Mixture-MCMC model. The results are summarized in Table \ref{tab:bayeisan_mcmc_result}. Despite having full knowledge of the underlying data generation processes for both the Wavelet and GARCH functions, the mixture model struggles to accurately approximate the posterior distribution in zero-shot forecasts, resulting in higher NLL and MSE compared to models trained on a single function type. This illustrates the phenomenon of negative transfer, where incorporating too many variations of data can negatively impact the modeling of the posterior distribution.

\textbf{Delphyne Training}
We observe similar effects while training our Delphyne model. During the pre-training phase, we evaluate different checkpoints to assess the zero-shot forecast performance on the ETTh2 dataset for two versions of our model: \methoda (trained with the LOTSA and financial data) and \methodl (trained without financial data).

We record the average Mean Absolute Error (MAE) across forecast lengths of $96$ and $196$. Additionally, we evaluate the fine-tuning performance of both models using MAE as the loss objective. Initially, \methoda incurs a higher MAE than \methodl, but after fine-tuning, both models achieve comparable MAE. For in-distribution forecasting on Monash dataset, we also observe that \methoda underperforms \methodl with larger aggregated MAEs. However, both methods perform comparably after fine-tuning (Sec. \ref{ssec:monash}).

\textbf{A Framework to Relate Pre-training to Downstream Performance} The three examples above demonstrate that negative transfer can indeed occur when pre-training time-series data from various domains. The continuous nature of time-series data, combined with its inherent noise and significant variability across domains, makes pre-training a model for time-series particularly challenging compared to other modalities like language and images.

\begin{wrapfigure}{r}{4cm}
\vspace{-0.35in}
\label{fig:ETTh2-zeroshot}
\includegraphics[width=0.3\textwidth]{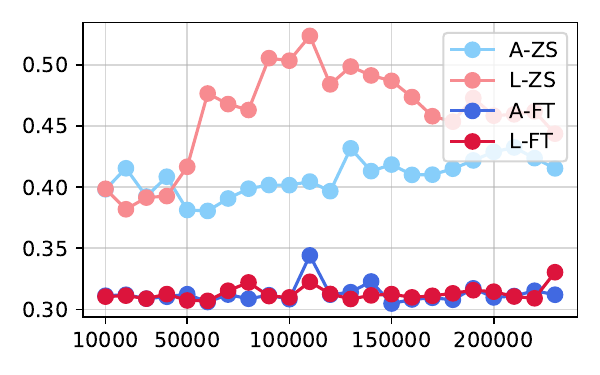}
\vspace{-0.33in}
\caption{MAE on ETTh2  across (average forecast length of $\{96,192\}$) training. Finance data hurts zero-shot (ZS) performance of  \methoda on ETTh2, but finetune (FT) eventually ``undoes'' the effect.}
\vspace{-0.25in}
\end{wrapfigure}

Given the negative transfer phenomenon, why do we need to build a pre-trained time-series model? The strength of a pre-trained time-series model lies in its ability to quickly adapt to downstream tasks with finetuning on only a few training samples, which offers great time and data efficiency in comparison to full shot models. 
We aim to develop a pre-trained time-series model that can swiftly adjust to new task distributions by efficiently “unlearning” pre-training biases and “adapting” to the specific characteristics of downstream tasks, even with limited training data or minimal gradient updates. In the next section, we detail our design choices and present ablation studies that underscore the critical role of these design elements in enabling the model to rapidly and effectively adapt during the fine-tuning stage.

\section{Model Overview}
\label{sec:method}

\textbf{Problem Formulation} For any-variate times-series, assuming that each dataset $\mathcal{D} = \{\mathbf{Y}^{(i)}\}_{i=1}^M$ has $M$ data points, while each $\mathbf{Y}^{(i)} \in \mathbb{R}^{l(i) \times T_{\mathbf{Y}{(i)}}}$, contains $l(i) (l(i) \geq 1)$ variates and $T_{\mathbf{Y}{(i)}}$ time-steps. 
We formulate the pre-training as a forecasting task: for each variate $j$, the future $h_j \geq 0$ time-steps are forecasted by modeling their forecast distribution: $P(\mathbf{Y}_{T_\mathbf{Y} - \mathbf{h}:T_{\mathbf{Y}}} | \phi)$, where $\phi$ is the output distribution of the time-series forecasting model and $\mathbf{h}$ is a vector comprised of $h_j$ time-steps. The overall training objective is: 
\begin{align}
\min_{\mathbf{w}} \text{      } \mathbb{E}_{ P(\mathbf{Y}) \sim \mathcal{D}}&\left[-\log P(\mathbf{Y}_{T_\mathbf{Y} - \mathbf{h}:T_{\mathbf{Y}}} | \phi)\right] \qquad
\text{s.t.              } \phi \leftarrow f_\mathbf{w}(\mathbf{Y}_{T_\mathbf{Y} - p :{T_\mathbf{Y} - \mathbf{h}}}
)
\end{align}
where we define $\mathbf{h}$ as the forecast length and $p$ as the look-back window length. Note that the number of time-steps conditioned on for each time series is different: $h_j - p$. We formulate our problem in this way where the number of time-steps to forecast is different per time series in order to allow for nowcasting where the information available for each time series can be different.

Time-series data, particularly in finance, exhibit unique characteristics that present distinct challenges: \begin{enumerate}
    \item \textbf{Multivariate Nature}: The variable $\mathbf{Y}$ typically has a dimensionality $k > 1$, indicating the presence of multiple interrelated time series. For example, US stocks are often quite correlated with the S\&P 500 index. 
    \item \textbf{Nowcasting Data}: The data of some variables is often available with considerable time lag, so it is important to estimate current value of the time-series based on it's own recent history and current values of other variables. This is the nowcasting task in economics and financial applications.
    \item \textbf{Multifrequency and Missing Data}: Each $\mathbf{Y}$ variable can be collected at varying granularities—such as individual tick data versus aggregated bar data—or may contain numerous missing entries due to irregular sampling intervals.
    \item \textbf{Extended Context Length}: Unlike natural languages, where input sequences are generally in the range of hundreds of tokens, financial time-series data often span thousands of timesteps, requiring models to handle significantly longer temporal dependencies.
\end{enumerate}

\subsection{Overall Architecture}
\label{ssec:overall_arch}

\begin{figure}
    \centering
    \vspace{-0.1in}
    \includegraphics[width=1.0\linewidth]{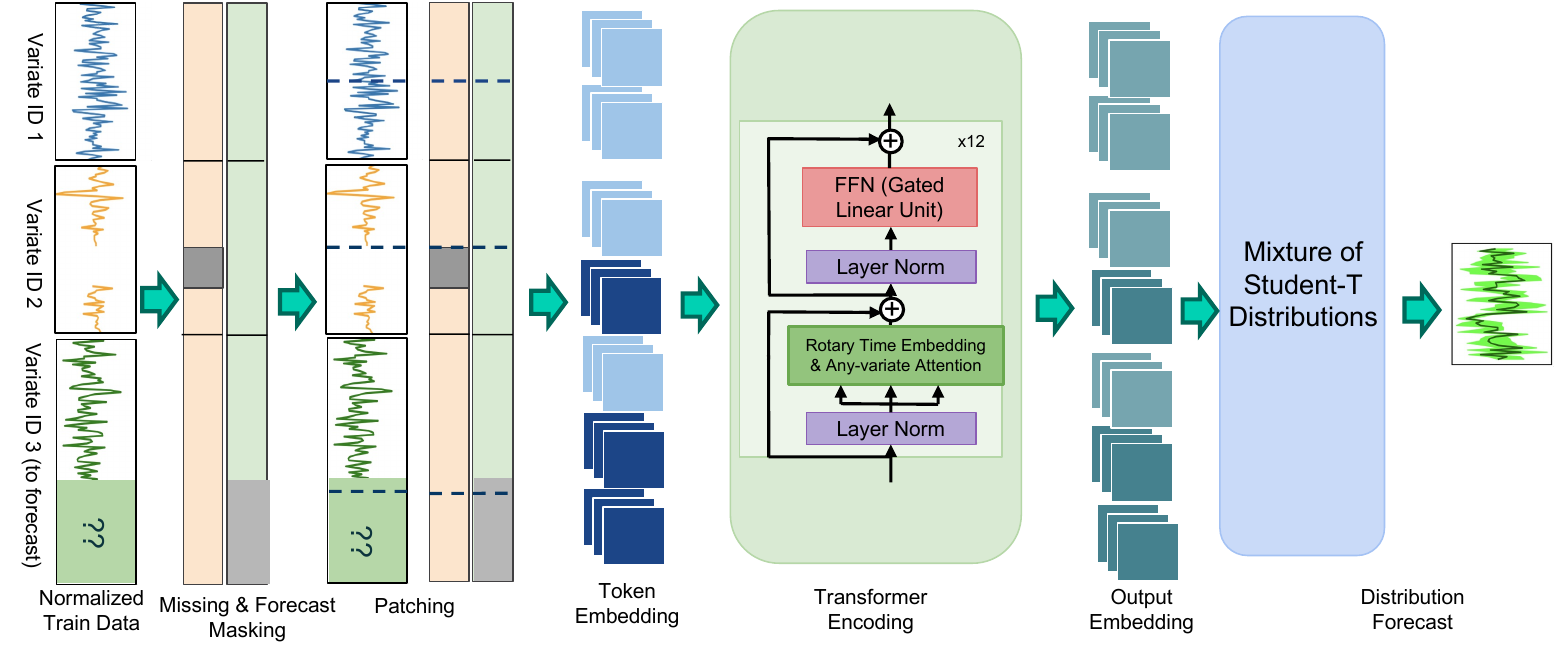}
    \vspace{-0.15in}
    \caption{Delphyne Overview}
\label{fig:overall_architecture}
    \vspace{-0.28in}
\end{figure}

Following recent approaches \citep{woo2024unifiedtraininguniversaltime,goswami2024moment}, we adopt a transformer encoder structure as the backbone of Delphyne. Fig. \ref{fig:overall_architecture} shows the overall pipeline.

\textbf{Padding.} To handle the multivariate nature of time-series data and accommodate the flexibility of varying the number of variates and covariates, we flatten and concatenate all variates along a single dimension. Each variate is assigned a unique identifier, and instance normalization is applied independently \citep{kim2021reversible}. For shorter time-series data, we pad the time series on the right (after forecast mask); prior literature pads to the left \cite{woo2024unifiedtraininguniversaltime,goswami2024moment}, but since we want to handle time series where each can have a variable-length, we must right-pad.

\textbf{Missing and Forecast Masking Before Patching.} Recent work shows that patching time series allows the model to attend to significantly longer contexts \citep{nie2023patchtst}. Therefore, Delphyne breaks the flattened time series into disjoint patches, fixing the patch size to $32$. Contrary to prior work, we create a \texttt{[FORECAST MASK]} to identify the target timesteps for forecast as well as a \texttt{[MISSING MASK]} that indicates where data is unobserved due to the sampling procedure. 
{For many finance tasks, data is collected at multiple frequencies or sampled irregularly, often leading to missing time steps (e.g., daily stock prices not being recorded on holidays). Simply ignoring these gaps can misalign multifrequency data, while backfilling or zero-filling distorts the original data distribution and can produce unsatisfactory forecast. We treat missing data similarly to forecast masks during training; however, we exclude the missing values from the forecast process. We apply both masks alongside the time-series data and learn a trainable linear projection to generate time-series embeddings that incorporate both data and missingness information. These patch embeddings serve as input to the transformer model. 
We suspect that applying masking before patching may lead to less effective token embeddings but is necessary for managing missingness in financial tasks. }

\textbf{Attention Block.} The transformer encoder utilizes pre-normalization \citep{xiong2020prenorm}, rotary positional embedding \citep{su2023roformerenhancedtransformerrotary}, any-variate attention (Sec. \ref{ssec:multivariate}), 
Silu activation function \citep{elfwing2017sigmoidweightedlinearunitsneural} and gated linear unit (GLU) \citep{shazeer2020gluvariantsimprovetransformer} to replace FFN.

\subsection{Context Length and Masking Ratio}
\label{ssec:context_length}
The Delphyne architecture requires careful tuning of key training parameters, like the masking ratio during pre-training. While vision transformers use around $60\%$ \citep{he2021maskedautoencodersscalablevision}, language models use $15\%$, and prior pre-trained time-series models range from $30\%$ to $50\%$ \citep{goswami2024moment,woo2024unifiedtraininguniversaltime}. This ratio is crucial as it affects input context length, impacting generalization and fine-tuning. We perform ablation studies to explore the impact of masking ratios and context lengths on training efficiency.

\textbf{Ablation 1.} The first experiment explores how varying pre-training context lengths affects both pre-training and fine-tuning performance. We train small- and medium-size autoregressive transformer decoders on synthetic Wavelet data with context lengths of $\{16,32,64,128\}$ (details in Appendix \ref{ssec:ablation_context_len}). After convergence, we assess zero-shot NLL forecast with a fixed context length of 32. Fine-tuning is performed with varying amounts of data, consistently using a context length of 32. Table \ref{tab:ablate_context_length} reports the NLL for zero-shot and fine-tuning. While pre-training with a context length of 32 yields the best zero-shot performance, a longer context length of $64,128$ improves fine-tuning, especially with fewer fine-tuning (10-100) samples. This suggests that for effective fine-tuning, pre-trained time-series models benefit from longer context lengths during pre-training.

\begin{table}[ht]
\centering
\caption{NLL($\downarrow$) for zero-shot and fine-tuning with varying sample sizes. Pretraining on a context length of 32 yields the best zero-shot performance, while pretraining on a longer context length of 128 improves fine-tuning, especially with 10-100 samples.}
\vspace{-0.13in}
\label{tab:ablate_context_length}
\resizebox{0.60\textwidth}{!}{%
\begin{tabular}{rccccc}
\toprule
Model & Zero-Shot & 1 Sample & 10 Samples & 100 Samples & 1000 Samples\\ \midrule
Small-16 &
-0.0449  &
-0.1189  &
-0.1523  &
-0.1824  &
-0.1870  \\
Small-32 & \textbf{-0.0978} & -0.0842 & -0.1681 & -0.1873 & -0.1898 \\
Small-64 &
-0.0808 &
\textbf{-0.1418} &
-0.1684 &
-0.1861 &
-0.1853
\\
Small-128 & -0.0842 & -0.1287 & \textbf{-0.1825} & \textbf{-0.1907} & \textbf{-0.1918} \\
\midrule
Medium-16 &
-0.0483 &
-0.0893 &
-0.1542 &
-0.1767 &
-0.1897\\
Medium-32 & \textbf{-0.1330} & \textbf{-0.1486} & -0.1673 & -0.1792 & -0.1847\\
Medium-64 &
-0.0793 &
-0.1146 &
\textbf{-0.1612} &
\textbf{-0.1873} &
-0.1875 \\
Medium-128 & -0.1020 & -0.1113& -0.1711 &  -0.1843 & \textbf{-0.1899} \\  \bottomrule
\end{tabular}}
\end{table}

\begin{wraptable}{r}{0.35\textwidth}
\centering
\caption{NLL($\downarrow$) for varying pre-training masking ratios.}
\vspace{-0.1in}
\label{tab:masking_ratio}
\resizebox{0.85\linewidth}{!}{%
\begin{tabular}{lcccc}
\toprule
{} & \textbf{Sample} & \multicolumn{3}{c}{\textbf{Masking Ratio} } \\ & \textbf{Size}& \textbf{0.25 } &\textbf{0.5} & \textbf{0.99} \\ \midrule
Pred. Len. 32 & 
\begin{tabular}[r]{c}1 \\ 10 \\ 100 \\ 1000\end{tabular} &
\begin{tabular}[c]{@{}l@{}} \textbf{-0.441} \\  \textbf{-0.394} \\ \textbf{-0.580} \\  -0.666 \end{tabular} & 
\begin{tabular}[c]{@{}l@{}}-0.187 \\  -0.071\\  -0.391 \\  -0.662 \end{tabular} &
\begin{tabular}[c]{@{}l@{}}0.416 \\  0.496\\ 0.321 \\ \textbf{-0.676} \end{tabular} \\ \midrule
Pred. Len. 64 & 
\begin{tabular}[r]{c}1 \\ 10 \\ 100 \\ 1000\end{tabular} &
\begin{tabular}[c]{@{}l@{}} \textbf{-0.263}\\  \textbf{-0.236} \\ \textbf{-0.296} \\  \textbf{-0.318} \end{tabular} & 
\begin{tabular}[c]{@{}l@{}}-0.077 \\  -0.077 \\  -0.158 \\  -0.314 \end{tabular} &
\begin{tabular}[c]{@{}l@{}}0.122 \\  0.144 \\  0.088 \\ -0.318 \end{tabular} \\ \midrule
Pred. Len. 96 & 
\begin{tabular}[r]{c}1 \\ 10 \\ 100 \\ 1000\end{tabular} &
\begin{tabular}[c]{@{}l@{}} \textbf{0.015} \\  \textbf{0.022} \\ \textbf{-0.006} \\  -0.020 \end{tabular} & 
\begin{tabular}[c]{@{}l@{}}0.081 \\  0.072 \\ 0.044 \\  \textbf{-0.023} \end{tabular} & 
\begin{tabular}[c]{@{}l@{}}0.132\\  0.144 \\  0.128 \\  -0.023 \end{tabular} \\ \bottomrule
\end{tabular}%
}
\vspace{-0.2in}
\end{wraptable}

\textbf{Ablation 2.} We vary the maximum masking ratio in $\{0.25, 0.5, 0.99\}$ and observe the effects of the resulting model. We use a transformer encoder and vary on {Wavelet} %
data (Appendix \ref{ssec:ablation_masking_ratio}). We evaluate the model's fine-tuning performance by NLL of various forecast length, see Table \ref{tab:masking_ratio}. When masking less aggressively, the model consistently performs well across various forecast lengths. This aligns with our earlier findings, as less aggressive masking implicitly exposes the model to longer context lengths during pre-training, leading to improved fine-tuning performance. 

During model pre-training, we apply a masking ratio to each variate, sampled from a beta-binomial distribution with $\alpha=5$ and $\beta = 10$, the average masking ratio is around $30\%$. Each variate is masked independently, enabling our model to better adapt to nowcasting scenarios where variates and covariates may have different lookback window lengths.

\subsection{Training on Multivariate Data}
\label{ssec:multivariate}
Multivariate time-series modeling poses unique challenges, particularly in capturing correlations between variates. \citet{nie2023patchtst} analyze channel-mixing versus channel-independence approaches, finding that the latter improves adaptability and generalization. Many pre-trained models for multivariate time series adopt channel-independence \citep{goswami2024moment,ekambaram2024tinytimemixersttms}, while others focus solely on univariate time series \citep{ansari2024chronos,das2024timesfm,rasul2024lagllamafoundationmodelsprobabilistic}. Recently, \citet{woo2024unifiedtraininguniversaltime} introduce any-variate attention (Appendix \ref{sec:any-variate_attention}
), which flattens multivariates into univariates and captures inter-variate correlations via a specialized bias term. We conduct experiments on synthetic Wavelet data to compare these design choices.

\textbf{Ablation 3. }We model two scenarios: one where the Wavelet data across rows are correlated, and another where they are uncorrelated. In the correlated scenario, the time-series data are generated using the same Wavelet function, differing only by additive Gaussian noise. In the uncorrelated scenario, the data are generated using different Wavelet functions, see Appendix \ref{ssec: multivariate} for details.

\begin{wraptable}{r}{0.6\textwidth}
\vspace{-0.1in}
\centering
\caption{NLL($\downarrow$) for zero-shot and fine-tuning with varying sample sizes, for different modeling multivariate methods.}
\vspace{-0.13in}
\label{tab:ablate_anyvariate}
\resizebox{0.59\textwidth}{!}{%
\begin{tabular}{rrccccc}
\toprule
\multicolumn{2}{r}{Model} & Zero-Shot & 1 Sample & 10 Samples & 100 Samples & 1000 Samples\\ \midrule
\multirow{3}{*}[-0.5pt]{\rotatebox[]{90}{Corr.}}  & Univariate & -0.0978 & -0.0842 & -0.1681  & -0.1873 & -0.1898 \\
 & Channel Mixing & \textbf{-0.1531} & \textbf{-0.1650} & \textbf{-0.1797} & -0.1873 & \textbf{-0.1913} \\
& Any-variate Attention & -0.1513 & -0.1507 & -0.1794 & \textbf{-0.1892} & -0.1895 \\ \midrule
\multirow{3}{*}[+1.8pt]{\rotatebox[]{90}{Uncorr.}} & Univariate & \textbf{-0.0978} & -0.0842 & -0.1681 &  -0.1873 &  \textbf{-0.1898} \\
& Channel Mixing & -0.0928 & -0.1323 & -0.1722 &  \textbf{-0.1892} &  -0.1874 \\
& Any-variate Attention & -0.0922 & \textbf{-0.1386}  & \textbf{-0.1879} &  -0.1879 & -0.1886 \\\bottomrule
\end{tabular}}
\vspace{-0.1in}
\end{wraptable}

Table \ref{tab:ablate_anyvariate} shows the results. Multivariate models significantly outperform univariate approaches for correlated rows of Wavelet data, underscoring the importance of capturing inter-variable dependencies.  Channel-mixing shows superior performance in zero-shot settings and with smaller training samples, but only when training and forecasting with correlated Wavelet data. However, when data rows are uncorrelated, channel-mixing performs worse than other two models. Given that Delphyne is trained on both {strongly and weakly} correlated financial data such as the returns of stocks {across various sectors}, it employs an any-variate attention mechanism. This mechanism enables the integration of cross-channel information without forcibly mixing variate data. Unlike  \cite{woo2024unifiedtraininguniversaltime}, we do not intentionally create multivariate data rows across different datasets.

\subsection{Output Distribution}
\label{ssec:output_dist}

For probability forecasting, previous studies usually assume a fixed output distribution \citep{salinas2019deeparprobabilisticforecastingautoregressive}. However, when the data presents varying distributions and supports, relying on a single distribution may be insufficient. 
\cite{woo2024unifiedtraininguniversaltime} propose a mixture of 
a Student’s T-distribution, Log-normal, Gaussian, and Negative Binomial, which introduces asymmetry into the forecasted distribution and demonstrates improved results. 
We conduct an additional experiment studying the effect of various distributions.  

\textbf{Ablation 4.} With a fixed training regime and transformer backbone, we measure a stock return's NLL (Sec. \ref{ssec:financial_tasks}) across three output distributions: a single distribution, a Student’s T mixture, and a mixture of different distributions as in \cite{woo2024unifiedtraininguniversaltime}, which we call MOIRAI’s Mixture (see details in Appendix \ref{ssec:output_distribution_ablation}). Fig. \ref{fig:distributions} shows that the Student’s T mixture performs similarly to MOIRAI’s Mixture on measuring stock returns. On the other hand, a single distribution underperforms both. For simplicity (Occam’s Razor), we choose the Student’s T mixture as the output distribution, with further discussion in the probability quantification experiment (Sec. \ref{ssec:probability_quant_exp}).

\begin{wrapfigure}{r}{0.3\linewidth}
    \centering
\includegraphics[width=\linewidth]{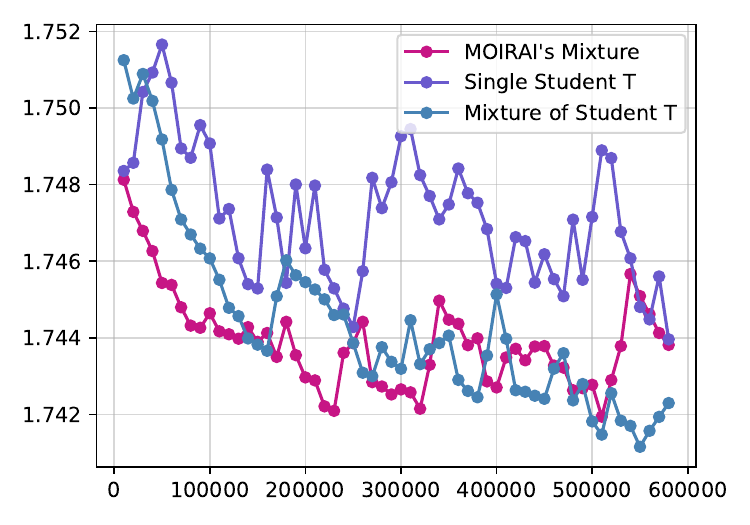}
    \vspace{-0.25in}
    \caption{Stock NLL performance on three different output distributions. }
\label{fig:distributions}
\end{wrapfigure}

\section{Training Details}
\label{sec:training}

\subsection{Training Data}
\label{ssec:training_data}
We include both publicly available data and  financial data in Delphyne's training, allowing Delphyne to generalize well to daily time-series forecast tasks as well as financial time-series tasks. We give a brief overview of our two sources of training data. 

\textbf{LOTSA Data} \citep{woo2024unifiedtraininguniversaltime} is the largest publicly available dataset for pre-training large time-series models, containing 231 billion observations across domains like energy, transport, climate, CloudOps, web, sales, nature, economics, finance, and healthcare. It aggregates most open source time-series datasets. An overview is provided in Appendix \ref{ssec:lotsa}.

\textbf{Financial Data} Our financial dataset contains data for companies, stocks, ETFs, currencies (exchange rates), and commodities. Further, to handle multiple frequencies, we create daily datasets and monthly datasets. To ensure that there is no lookahead bias in our downstream tasks, we pre-train with data only until the end of 2019. 
We provide a detailed breakdown of the dataset in Appendix \ref{ssec:fin_data}.

\textbf{Sampling}
The LOTSA dataset is significantly imbalanced, necessitating subsampling to ensure more balanced representation during training. We carefully identify the few datasets that dominate in size and reduce their likelihood of being sampled to avoid overrepresentation. For any dataset, we first compute the total number of observations (across samples, variates, and timesteps) within the dataset, $|\mathcal{D}_k| = \sum_{i=1}^M\sum_{j=1}^{l+k} T_{i,j}$. 
Then, we normalize the scores to sum to one.
We cap the weighting to be at least $0.001$ which we normalize again to obtain the final probability of sampling a dataset. Overall, our training data consists of 85\% from LOTSA and 15\% from financial data (Table \ref{tab:prob_sampling}). We sample the number of variates ($\leq 128$) using a beta-binomial distribution ($\alpha=2$ and $\beta=5$).

\begin{table}[htbp]
    \centering
    \vspace{-0.05in}
    \caption{Sampling dataset probability (\%) across LOTSA domains and finance data.}
    \label{tab:prob_sampling}
    \vspace{-0.1in}
    \resizebox{0.9\textwidth}{!}{
        \begin{tabular}{lcccccccccc}
            \toprule
            \textbf{LOTSA} & Energy & Transport & CloudOps & Climate & Econ/fin & Web & Sales & Nature & Healthcare & \textbf{Total} \\ 
             & 17.8 & 25.5 & 8.6 & 11.9 & 5.7 & 6.1 &  5.4 & 3.9 & 0.2 & \textbf{85} \\ 
             \midrule
            \textbf{Finance Data} & & ETFs & Tickers & Commodities & Currency & Stock & Company & Intraday Bars  &   & \textbf{Total} \\ 
             &  & 2.8 & 2.8 & 0.3 &  0.9 & 2.8 &  1.8 & 2.8 &  & \textbf{15} \\  
            \bottomrule
        \end{tabular}
        \vspace{-0.35in}
    }
\end{table}

Unlike other models \citep{woo2024unifiedtraininguniversaltime,goswami2024moment} that involve complex preprocessing like packing, concatenation, or augmentation, our approach is streamlined. We only apply subsampling and truncate time to $512 \times 32$ timesteps. For the ease of normalization, we exclude entries with fewer than five observations or uniform values after forecast and missing masking. This keeps the training data robust and optimizes model performance without excessive preprocessing.

\subsection{Model Parameters}

We train Delphyne with $12$ layers and $768$-dimensional attention shared across $12$ heads, with a maximum width of $3072$. Dropout is set to $0.2$, and the model is trained on negative log-likelihood (NLL) loss. We pretrained for 1 million gradient updates with a fixed patch size of $32$ and a sequence length of $512 \times 32$ steps. Using a batch size of 256, we optimize with AdamW (learning rate = $1e-4$, weight decay = $0.1$, $\beta_1 = 0.9$, $\beta_2 = 0.98$) and apply a learning rate scheduler with $10,000$ steps of linear warmup and cosine annealing to $1e-5$. Training was conducted on 8 H100 GPUs over ~4 days
with mixed-precision (bf16) \citep{Kalamkar2019ASO} for speed.

\section{Experiments}
\label{sec:exp}
We train three models for downstream evaluations. Delphyne trained on the LOTSA dataset \citep{woo2024unifiedtraininguniversaltime}, Delphyne trained on finance data only, and Delphyne trained with all finance and LOTSA dataset, which we define as \methodl, \methodf, and \methoda, respectively. We compare the zero-shot (ZS) forecasts and fine-tuning (FT) performances on standard short-term and long-term time-series forecasting tasks, probabilistic quantification, anomaly detection tasks, as well as the finance forecast tasks for stock risk analysis, stock volatility modeling, intraday traded volume modeling, and company revenue nowcasting.

\subsection{Financial Tasks}
\label{ssec:financial_tasks}
\textbf{Stocks.} While a commonly evaluated task for finance is forecasting the returns, modeling the distribution of stock returns is equally important where it can be used for stress-testing scenarios, and conducting risk analysis \citep{tepelyan2023equitygenerativemodeling}.
To test the performance of different methods on this task, we use the daily stock returns of 14 major stocks from SPX Index from 2021-01-04 to 2023-12-29. For fine-tuning, we use data from 1996-07-01 to 2019-12-31 for training, 2020-01-02 to 2020-12-31 for validation with a context length of 252. 

Since most methods, except MOIRAI \citep{woo2024unifiedtraininguniversaltime}, forecast point estimates, we conduct two experiments: forecasting next-day stock returns' variance (volatility) to compare MSE, and evaluating NLL (Negative Log Likelihood) for the forecasted returns distribution. We also benchmark against GARCH with Student’s T, a standard financial baseline \citep{tepelyan2023equitygenerativemodeling}. For PatchTST \citep{nie2023patchtst}, we adapt the output to a mixture of four Student’s T distributions.

Table \ref{tab:stock_nll} and \ref{tab:stock_mse} show the overall results.The NLL metrics indicate that while utilizing only financial data in pre-training brings best zero-shot performance (\methodf), \methoda achieves the best results after fine-tuning. The MSE for variance forecasts shows that Delphyne models outperform others and perform better in zero-shot than after fine-tuning. We hypothesize that the dataset's high noise leads to overfitting during fine-tuning. Additional evaluation of coverage statistics (a measure of the model's calibration performance) confirms that Delphyne models achieve the best performance (Appendix \ref{ssec:cov_stocks}).

\begin{table}[htbp]
\centering
\begin{minipage}[t]{0.23\textwidth}
    \centering
    \caption{Likelihood results for next-day stock returns risk analysis.}
    \vspace{-0.1in}
    \label{tab:stock_nll}
    \scalebox{0.6}{
    \begin{tabular}{lcc}
    \toprule
    \textbf{Model}      & \textbf{NLL ZS} & \textbf{NLL FT} \\
    \midrule
    \textbf{\methoda}    & 1.762          & \textbf{1.741}  \\
    \textbf{\methodf}    & 1.750          & \underline{1.746}  \\
    \textbf{\methodl}    & 1.775          & 1.757  \\
    \textbf{MOIRAI}      & 1.776          & 1.788  \\
    \textbf{GARCH}       & -              & 1.752  \\
    \textbf{PatchTST}    & -              & {1.751} \\
    \bottomrule
    \end{tabular}
    }
\end{minipage}
\hfill
\begin{minipage}[t]{0.23\textwidth}
    \centering
    \caption{MSE for next-day stock squared returns (variance).}
    \vspace{-0.09in}
    \label{tab:stock_mse}
    \scalebox{0.6}{
    \begin{tabular}{lcc}
    \toprule
    \textbf{Model}      & \textbf{MSE ZS} & \textbf{MSE FT} \\
    \midrule
    \textbf{\methoda}    & 37.792          & 37.810  \\
    \textbf{\methodf}    & \underline{37.653}          & 38.616  \\
    \textbf{\methodl}    & \textbf{37.591}          & 38.246  \\
    \textbf{MOIRAI}      & 41.428          & 40.502  \\
    \textbf{MOMENT}      & 46.006          & 37.785  \\
    \textbf{TTM}         & 44.918          & 44.360  \\
    \textbf{PatchTST}    & -               & 51.705  \\
    \textbf{GARCH}       & -               & 41.517  \\
    \bottomrule
    \end{tabular}
    }
\end{minipage}
\hfill
\begin{minipage}[t]{0.23\textwidth}
    \centering
    \caption{MSE for bars log-volume data. (78 timestep predictions of 5-minute intervals)}
    \label{tab:bars}
    \vspace{-0.1in}
    \scalebox{0.55}{
    \begin{tabular}{lcc}
    
\toprule
\textbf{Model}    & \textbf{MSE ZS} & \textbf{MSE FT} \\
\hline
\textbf{\methoda} & 0.728   & 0.551      \\
\textbf{\methodf} & 0.965  & \textbf{0.530}       \\
\textbf{\methodl} & 0.930   & 0.557      \\
\textbf{MOIRAI}      & 0.767    & 0.620     \\
\textbf{MOMENT}      & 0.775     & 0.838  \\
\textbf{TTM}         & 0.714     & 0.601    \\
\textbf{PatchTST}    &    -    & \underline{0.534}  \\
\textbf{Avg past values}  & 0.602   & -   \\
\bottomrule
\end{tabular}
}
\end{minipage}
\hfill
\begin{minipage}[t]{0.23\textwidth}
    \centering
    \caption{Nowcasting results for zero-shot vs. fine-tuning for company sales growth}
    \label{tab:bsm}
    \scalebox{0.6}{
    \begin{tabular}{lcc}
    \toprule
    \textbf{Model}      & \textbf{MAE ZS} & \textbf{MAE FT} \\
    \midrule
    \textbf{\methoda}    & 0.099         & \textbf{0.071} \\
    \textbf{\methodf}    & 0.128       & 0.079  \\
    \textbf{\methodl}    & 0.101        & \underline{0.073}  \\
    \textbf{MOIRAI}      & 0.091    & 0.093     \\
    \textbf{Baseline}      & 0.100    & -     \\
    \bottomrule
    \end{tabular}
    }
\end{minipage}
\end{table}

\textbf{Bars.}   We use the intraday bars data, which contains the log of volume traded (log-volume) in five-minute intervals to test the different methods' performance in long-sequence modeling. For 4 different ETFs, we use the past 15 days data (context length $15 \times 78$) to forecast the log-volume in the next day's trading hours (e.g., forecast length of 78) and compare their mean-squared errors.

Table \ref{tab:bars} shows the results for 2021-01-04 to 2021-01-11. For fine-tuning, data from 2008-01-24 to 2019-12-30 is used for training, and 2019-12-31 to 2020-12-31 for validation. All Delphyne models significantly improve metrics after fine-tuning, effectively capturing the seasonal component in bar log-volume data. \methoda-ZS outperforms \methodf-ZS and \methodl-ZS due to the data's seasonal nature, similar to electricity and weather datasets. However, with fine-tuning, \methodf achieves the best performance across all methods.

\textbf{Nowcasting Company Revenue.} We use consumer transaction data to test nowcasting performance (e.g., when we have contemporaneous data), forecasting year-over-year (YoY) sales growth for 211 U.S. companies based on 1) previous quarter's YoY sales growth, 2) previous YoY growth in transactions, and 3) current quarter's YoY growth in transactions. Due to the quarterly nature, the context length is short (4-8). We make rolling forecasts for Q3 2022 to Q1 2023. For fine-tuning, data from 2018 Q1 to 2021 Q1 is used for training, and 2021 Q2 to 2022 Q2 for validation. We compare against a statistical baseline built by the providers of the consumer transaction data and MOIRAI \citep{woo2024unifiedtraininguniversaltime}. Table \ref{tab:bsm} shows MSE results: \methoda with fine-tuning outperforms all methods, and \methodl ranks second despite not seeing the data during pre-training. We attribute this to \methodl's independent masking of variates during pre-training, which enhances handling of contemporaneous data for nowcasting.

\subsection{Short-term Forecasting on Monash Dataset}
\label{ssec:monash}
We conduct an evaluation using the Monash dataset \citep{godahewa2021monash}, which spans multiple domains like demand forecasting, traffic, and weather, with various data granularities. We follow the train-test split outlined in \cite{woo2024unifiedtraininguniversaltime}, evaluating performance only on the hold-out test set to ensure a fair in-distribution comparison for \methodl and \methoda, whose pre-training datasets contain training sets of Monash data.
We partition out the same forecast length as the test set, used as validation data for finetuning the model. We report both zero-shot and finetuning results. For zero-shot, we fix context length equals $500$ and report the test performance. For fine-tuning, we search context length in $\{100, 300,500, 1000, 2000\}$ and the learning rate in $\{5e-5,1e-5\}$.

\begin{figure}[htbp]
\vspace{-0.15in}
    \begin{minipage}[t]{0.6\linewidth} %
        \includegraphics[width=\linewidth]{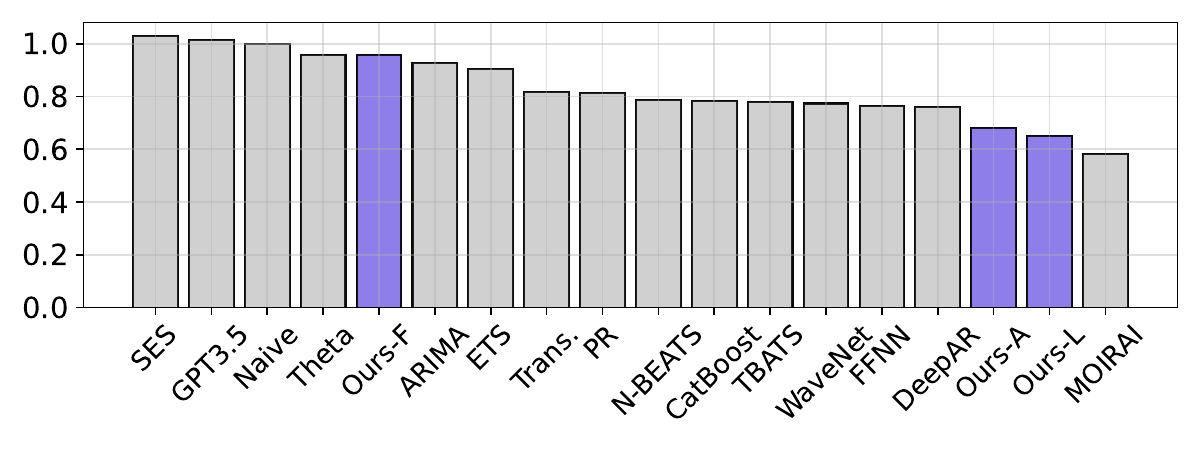}
        \vspace{-0.3 in}
        \caption{Aggregated geometric mean of normalized MAEs on the Monash Time-Series Forecasting Benchmark. On average, Delphyne zero-shot models perform better than existing models, falling into second place behind MOIRAI.}
        \label{fig:monash_zeroshot}
    \end{minipage}
    \hfill
    \begin{minipage}[t]{0.35\linewidth} %
    \includegraphics[width=0.98\linewidth]{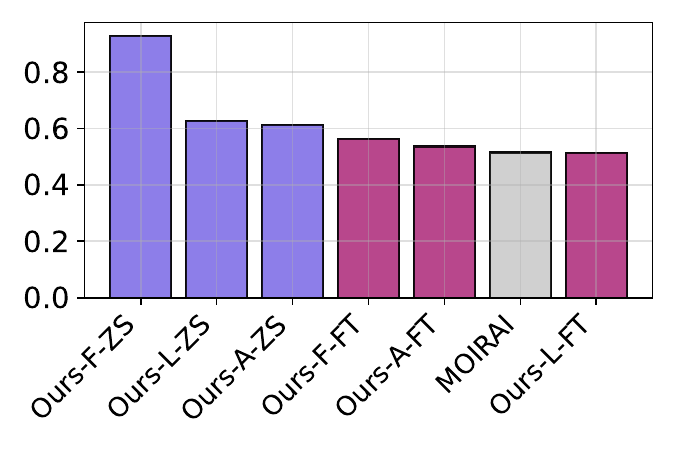}
         \vspace{-0.15in}
        \caption{Aggregated geometric mean of normalized MAEs for Delphyne zero-shot (ZS) and fine-tuned(FT) models. }
        \label{fig:monash_ft}
    \end{minipage}
\end{figure}

We compare Delphyne with statistical baselines, supervised ML baselines, foundation models, and LLM-based models. Full MAE result for comparison is in Table \ref{tab:monash_results} and comparison across versions of Delphyne is in Table \ref{tab:monash_delphyne}. Fig. \ref{fig:monash_zeroshot} and Fig. \ref{fig:monash_ft} provide the aggregate results of normalized MAE (the geometric mean of the MAE of each dataset divided by the MAE of the Naive approach).

In zero-shot performance, \methodf, trained solely on finance data, slightly outperforms the Naive model and the other three benchmarks. Since Monash is out-of-distribution for \methodf, a decline in performance is expected. In contrast, \methodl achieves a lower MAE than \methoda, likely because that \methoda's training data contains finance data and may introduce negative transfer effects. After fine-tuning, both \methoda and \methodl show similar performance, rivaling the best model, MOIRAI \citep{woo2024unifiedtraininguniversaltime}.

\subsection{Out-of-distribution Long-term Forecasting}
For long-term forecasting, we evaluate Delphyne and benchmarks on popular datasets with forecast lengths of ${96,192,336}$, and ${720}$. We use the standard ETT datasets, weather \citep{zhou2021informerefficienttransformerlong} and electricity(ECL) dataset \citep{electricityloaddiagrams20112014_321} . See Appendix \ref{sec:long-term-forecasting-experiment-details} for experiment details.

We report the average performance across various forecast lengths for zero-shot, linear probing, fine-tuning, and full-shot methods in Table \ref{tab:ett_agg}, with the full results provided in Table \ref{tab:ett_overall}. Admittedly, \methoda underperforms compared to TTM, MOIRAI, and other foundation models in zero-shot settings. This may be attributed to its use of fixed patch sizes rather than dynamic adaptation and Delphyne's training on the negative log-likelihood of a Student's-T mixture distribution, which is optimized for financial tasks. However, \methoda demonstrates strong performance after fine-tuning, a crucial step for improving out-of-distribution forecasting. Its architectural design and training paradigm make it particularly adaptable.

We provide a comparison of Delphyne models in Table \ref{tab:ett_zs_ft}
. \methoda significantly outperforms \methodl on ETTm1 and ETTh1 with a 720-forecast length, with no observed negative transfer. This may be due to the shared seasonality trends between ETT and financial (bars) data. However, both models achieve similar performance after fine-tuning.

\begin{table}[htbp]
\centering
\caption{Average MAE and MSE across prediction lengths $\{96,192,336,720\}$ for Delphyne and other baseline methods. The best are highlighted in \textbf{bold} and the runner-up is \underline{underlined}.}
\vspace{-0.1in}
\resizebox{\textwidth}{!}{
\begin{tabular}{lccccccccccccccccccccccccccccccccccc}
\toprule
 &\multicolumn{4}{c}{Zero-shot} &\multicolumn{2}{c}{Linear-Probing} &\multicolumn{1}{c}{Fine-tuned} &\multicolumn{5}{c}{Full-Shot} \\
 \cmidrule(lr){2-5} \cmidrule(lr){6-7} \cmidrule(lr){8-8}
\cmidrule(lr){9-13}
&\textbf{TTM} & \textbf{MOIRAI} & \textbf{TimesFM} & \methoda-ZS  & \textbf{MOMENT} & \textbf{GPT4TS} & \methoda-FT &  \textbf{PatchTST} & \textbf{Dlinear} & \textbf{TimesNet} & \textbf{FEDFormer} & \textbf{Stationary}  \\
\midrule
\textbf{ETTh1}&\\
MSE & \textbf{0.402} & 0.434 & 0.476  & 0.449 & 0.418 & 0.428 & 0.440 & \underline{0.413} & 0.423 & 0.457 & 0.440 & 0.570 \\
MAE & - & 0.439 & 0.451 & 0.450 & 0.436 & \textbf{0.426} & 0.441 & \underline{0.431} & 0.437 & 0.449 & 0.460 & 0.537\\
\textbf{ETTh2}& \\
MSE & \textbf{0.327} & \underline{0.346} & 0.404 & 0.375 & 0.352 & 0.355 & 0.352 & 0.330 & 0.431 & 0.414 & 0.437 & 0.526 \\
MAE &- & 0.382 & 0.406 & 0.404 & 0.395 & 0.395 & \textbf{0.356} & \underline{0.379} & 0.447 & 0.427 & 0.449 & 0.516 \\
 \textbf{ETTm1}& \\
 MSE & \textbf{0.338} & 0.382 & 0.420 & 0.501 & 0.354 & 0.352 & 0.364 & \underline{0.351} & 0.357 & 0.400 & 0.448 & 0.481\\
 MAE & - & 0.388 & 0.408 & 0.429 & 0.391 & 0.383 & \textbf{0.365} & 0.381 & \underline{0.379} & 0.406 & 0.452 & 0.456\\
 \textbf{ETTm2}& \\
 MSE & 0.264 & 0.272 & 0.350 & 0.323 & 0.256 & 0.266 & \textbf{0.250} & \underline{0.255} & 0.267 & 0.291 & 0.305 & 0.306\\
MAE & - & 0.321 & 0.353 & 0.363 & \underline{0.270} & 0.326 & \textbf{0.257} & 0.315 & 0.334 & 0.333 & 0.350 & 0.347 \\
 \textbf{ECL}& \\
 MSE & \underline{0.160} & 0.188 & \textbf{0.156} & 0.202 & 0.165 & 0.167 & 0.170 & 0.162 & 0.166 & 0.193 & 0.214 & 0.193\\
 MAE & - & 0.274 & \underline{0.246} & 0.293 & 0.260 & 0.263 & \textbf{0.203} & 0.253 & 0.264 & 0.295 & 0.327 & 0.296\\
 \textbf{Weather}& \\
 MSE & 0.233 & 0.235 & 0.232 & 0.369 & 0.230 & 0.237 & \textbf{0.221} & \underline{0.226} & 0.249 & 0.259 & 0.309 & 0.288\\
 MAE & - & 0.263 & \underline{0.257} & 0.348 & 0.261 & 0.271 & \textbf{0.235} & 0.264 & 0.300 & 0.286 & 0.360 & 0.314\\
\bottomrule
\end{tabular}}
\label{tab:ett_agg}
\vspace{-0.15in}
\end{table}

\subsection{Probability Quantification}
\label{ssec:probability_quant_exp}
We assess probability forecasting on six datasets spanning the energy, transport, climate, and sales domains, using a rolling evaluation setup where the stride matches the forecast length. Additional experiment setups, comparison methods, and full results are shown in Appendix \ref{sec:prob_forecast_appendix}.

We report the Continuous Ranked Probability Score (CRPS)
and Mean Scaled Interval Score (MSIS) metrics (definitions
in Appendix \ref{sec:prob_forecast_appendix}) in Table \ref{tab:prob_forecast_short}, and additional deterministic metrics are shown in Table \ref{tab:prob_forecast_full}.
\methoda's zero-shot performance is slightly below that of MOIRAI \citep{woo2024unifiedtraininguniversaltime}, particularly in terms of CRPS metrics. However, after fine-tuning, \methoda shows significant improvement, achieving the best results across various datasets. We suspect that we outperform in MSIS likely due to \methoda modeling having a heavier-tailed distribution using a mixture of Student's T distributions, but there is still room in selecting a more expressive distribution to improve CRPS, as distributions can be specific to each dataset.

\begin{table}[htbp]
\centering
\caption{Full results for probabilistic forecasting experiments. The best results are highlighted in \textbf{bold}, and the second best results are \underline{underlined}. (The baseline results are taken from \cite{woo2024unifiedtraininguniversaltime}.)}
\vspace{-0.12 in}
\label{tab:prob_forecast_short}
\resizebox{1.0\textwidth}{!}{%
\begin{tabular}{lccccccccccc}
\toprule
& & \multicolumn{2}{c}{\textbf{Zero-shot}} & \multicolumn{1}{c}{\textbf{Finetuned}} & \multicolumn{4}{c}{\textbf{Full-shot}} & \multicolumn{2}{c}{\textbf{Baseline}}\\ 
\cmidrule(lr){3-4} \cmidrule(lr){5-5} \cmidrule(lr){6-9} \cmidrule(lr){10-11}
& & \methoda-ZS & MOIRAI & \methoda-FT &  PatchTST  & TiDE & TFT & DeepAR & AutoARIMA & Seasonal Naive   \\
\midrule
\multirow{2}{*}{\textbf{Electricity}} 
& \textbf{CRPS} &  0.159 & 0.055 & 0.140$\pm$0.005 &  0.052$\pm$0.00 & \textbf{0.048$\pm$0.00} & \underline{0.050$\pm$0.00} & 0.065$\pm$0.01 & 0.327 & 0.070 \\
& \textbf{MSIS} & 29.293 & 6.172 & 21.820$\pm$1.383 & \underline{5.744$\pm$0.12} & \textbf{5.672$\pm$0.08} & 6.278$\pm$0.24 & 6.893$\pm$0.82 & 29.412 & 35.251 \\ 
\midrule
\multirow{2}{*}{\textbf{Solar}} 
& \textbf{CRPS} & 0.905 & 0.419 & 1.306$\pm$0.103 & 0.518$\pm$0.09 & \textbf{0.420$\pm$0.00} & 0.446$\pm$0.03 & \underline{0.431$\pm$0.01} & 1.055 & 0.512 \\
& \textbf{MSIS} & \underline{2.733} & 7.011 & \textbf{2.029$\pm$0.520}  & 8.447$\pm$1.59 & 13.754$\pm$0.32 & 8.057$\pm$3.51 & 11.181$\pm$0.67 & 25.849 & 48.130 \\
\midrule
\multirow{2}{*}{\textbf{Walmart}} 
& \textbf{CRPS} &  0.093 & 0.093 & 0.083$\pm$0.001 &  \underline{0.082$\pm$0.01} & \textbf{0.077$\pm$0.00} & 0.087$\pm$0.00 & 0.121$\pm$0.00 & 0.124 & 0.151 \\
& \textbf{MSIS} & \underline{4.741} & 8.421 & \textbf{4.559$\pm$0.289} & 6.005$\pm$0.21 & 6.258$\pm$0.12 & 8.718$\pm$0.10 & 12.502$\pm$0.03 & 9.888 & 49.458 \\
\midrule
\multirow{2}{*}{\textbf{Weather}} 
& \textbf{CRPS} & 0.064 & \textbf{0.041} & \underline{0.042$\pm$0.005} & 0.059$\pm$0.01 & 0.054$\pm$0.00 & 0.043$\pm$0.00 & 0.132$\pm$0.01 & 0.252 & 0.068 \\
& \textbf{MSIS} & 6.080 &  \underline{5.136} & \textbf{4.467$\pm$0.053} & 7.759$\pm$0.49 & 8.095$\pm$1.74 & 7.791$\pm$0.44 & 21.651$\pm$17.34 & 19.805 & 31.293 \\
\midrule
\multirow{2}{*}{\textbf{Istanbul Traffic}} 
& \textbf{CRPS} & 0.149 & 0.116& 0.212$\pm$0.015 & 0.112$\pm$0.00 & 0.110$\pm$0.01 & \underline{0.110$\pm$0.01} & \textbf{0.108$\pm$0.00} & 0.589 & 0.257 \\
& \textbf{MSIS} & 9.989 & 4.461 & 4.328$\pm$0.536 & \textbf{3.813$\pm$0.09} & 4.752$\pm$0.17 & \underline{4.057$\pm$0.44} & 4.094$\pm$0.31 & 16.317 & 45.473 \\
\midrule
\multirow{2}{*}{\textbf{Turkey Power}} 
& \textbf{CRPS} & 0.046 & \underline{0.040} & \textbf{0.035$\pm$0.001} & 0.054$\pm$0.01 & 0.046$\pm$0.01 & 0.039$\pm$0.00 & 0.066$\pm$0.02 & 0.116 & 0.085 \\
& \textbf{MSIS} & \underline{6.269} & 6.766 & \textbf{5.384$\pm$0.346} & 8.978$\pm$0.51 & 8.579$\pm$0.52 & 7.943$\pm$0.31 & 13.520$\pm$1.17 & 14.863 & 36.256 \\
\bottomrule
\end{tabular}%
}
\end{table}
 
\vspace{-0.2in}
\subsection{Anomaly Detection}
\vspace{-0.09in}

\begin{wrapfigure}{r}{0.30\textwidth}
    \vspace{-0.35in}
    \centering    \includegraphics[width=0.30\textwidth]{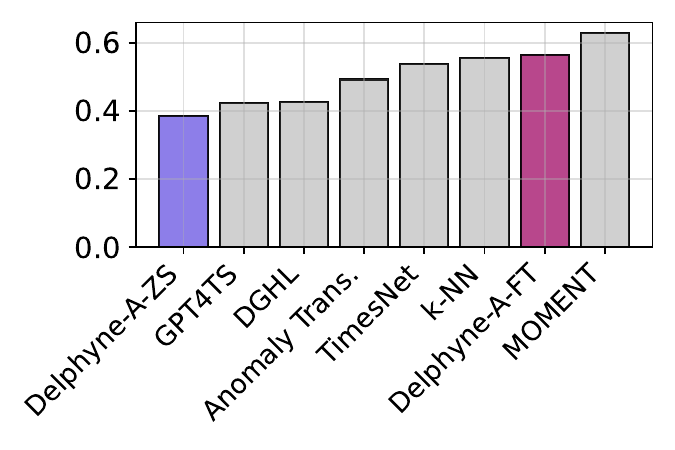}
    \vspace{-0.3in}
    \caption{Aggregated Adjusted F1 Score for \methoda vs. comparison baselines.}
    \label{fig:anomalydetection}
    \vspace{-0.2in}
\end{wrapfigure}

We measure adjusted F1 score for the anomaly detection task, on 44 time-series datasets for the UCR anomaly detection archive, in comparison to popular full-shot models and foundation model MOMENT with anomaly detection head \citep{goswami2024moment}. Experiment setups and full results are in Appendix \ref{sec:anomaly_detection_description}.

The aggregated F1 score is provided in Fig. \ref{fig:anomalydetection}.  \methoda's versatility allows it to adapt well to anomaly detection tasks after fine-tuning, achieving second place overall in anomaly detection tasks.

\section{Related Work}
\label{sec:related}
\textbf{Transformer-based Time-series Modeling.} Transformers have become a powerful backbone for time-series analysis \citep{wen2023transformerstimeseriessurvey}. Many studies enhance attention mechanisms to handle longer context lengths in time-series data \citep{zhou2021informerefficienttransformerlong,zhou2022fedformerfrequencyenhanceddecomposed}. \citet{nie2023patchtst} show that segmenting time-series data into patches as tokens improves performance and captures representations effectively for forecasting. Similarly, \citet{liu2024itransformerinvertedtransformerseffective} treat independent time series as tokens to capture inter-correlations, enhancing multivariate forecasting.

\textbf{Time-series Foundation Models.} Building on the rise of  LLMs, recent research has explored the potential of large deep learning models for time-series analysis. One approach adapts existing pre-trained language models for time-series tasks through prompting \citep{zhou2023ofa,jin2024timellm,chang2024llm4ts,gruver2024LLMTime,Liu2024LSTPromptLL}, while another focuses on training foundation models specifically on time-series data \citep{woo2024unifiedtraininguniversaltime,das2024timesfm,goswami2024moment,ekambaram2024tinytimemixersttms}. Both strategies seek to develop a single model trained on large-scale, cross-domain data, capable of addressing multiple forecasting tasks, in contrast to full-shot methods that require training a separate model for each dataset. Our work is closely related to foundation models (See Appendix \ref{sec:list_of_popular_models}), with several architectural modifications for financial domains.

\textbf{Transfer Learning and Distribution Shift in Time-series domains.}
The negative transfer effect is widely studied in the transfer learning literature, but remains less explored in the context of pre-training models for time-series modeling. This effect can be viewed through the lens of distribution shifts within training and testing time-series datasets. Several studies have proposed methods to address distribution shifts in time-series tasks \citep{fan2023dishtsgeneralparadigmalleviating,kim2021reversible,liu2024timeseriesforecastingoutofdistributiongeneralization,Liu2024TimeMMDAN}. Our work differs from the distribution shift literature; we aim to develop a pre-trained time-series  model capable of rapidly adapting to downstream tasks through few-shot fine-tuning.

\section{Conclusions}
\label{sec:conclusions}
We illustrate the presence of negative transfer effect in pre-trained time-series models especially when pre-trained with time-series data from various domains, contrasting them with LLMs for language tasks. Our experiments emphasize the role of fine-tuning to counter this effect which helps pre-trained time-series models to adapt to diverse downstream tasks with few training samples and minimal iterations. We introduce various architectural modifications, supported by ablation studies, to handle continuous, noisy, multivariate and multifrequency nature of time-series data. Delphyne is the first pre-trained model excelling in both general time-series and diverse financial and economic tasks such as nowcasting.

\bibliographystyle{main}
\bibliography{ref}

\appendix
\newpage

\section{Any-variate Attention}
Any-variate attention is first proposed by 
\label{sec:any-variate_attention}
\cite{woo2024unifiedtraininguniversaltime} to allow binary attention biases to encode variate indices for a flattened multi-variate time series. The attention score between the $(i, m)$-th query and $(j,n)$-th query ($j$ and $i$ represent the time-steps, and $n$ and $m$ encode the variate index) is calculated as the following:

\begin{align}
E_{ij,mn} &= \left( \mathbf{W}^Q \mathbf{x}_{i,m} \right)^T \mathbf{R}_{i-j} \left( \mathbf{W}^K \mathbf{x}_{j,n} \right)
+ u^{(1)} * \mathbbm{1}_{\{m=n\}} + u^{(2)} * \mathbbm{1}_{\{m \neq n\}}, \\
A_{ij,mn} &= \frac{\exp(E_{ij,mn})}{\sum_{k,o} \exp(E_{ik,mo})},
\end{align}

where \(\mathbf{W}^Q \mathbf{x}_{i,m}, \mathbf{W}^K \mathbf{x}_{j,n} \in \mathbb{R}^{d_h}\) are the query and key vectors. $u^{(1)}$ and $u^{(2)}$ are learnable scalars as the attention biases. These binary attention biases component enables differentiation between variates,  satisfies permutation equivariance and invariance with respect to variate ordering, and is scalable to any number of variates.

\section{Pre-training Data}

\subsection{LOTSA}\label{ssec:lotsa}

\begin{enumerate}
\item \textbf{BuildingsBench} BuildingsBench \citep{emami2023buildingsbench} comprises of datasets detailing energy consumption in residential and commercial buildings. These include the real-world BDG-2 datasets, Low Carbon London, SMART, IDEAL, Sceaux, and Borealis, which capture energy usage from diverse sources. BuildingsBench introduces the Buildings-900K dataset, a large-scale simulation of 900K buildings, while both the training and testing splits are included in LOTSA. Electricity is omitted in LOTSA and used for out-of-distribution evaluation.
\item \textbf{ClimateLearn} This dataset includes both ERA5 and CMIP6 \citep{nie2023patchtst}, which contain various climate-related variables like temperature and humidity across different pressure levels. In LOTSA, we observe that ERA5 and CMIP6 are divided into several data folders across different years. To address this, we reduce the probability of their appearance by treating all directories spanning multiple years as single datasets.
\item \textbf{CloudOps} CloudOps-TSF, introduced by \cite{woo2024pushing}, provides three large-scale time series datasets that capture variables like CPU and memory utilization. Only training dataset is included in LOTSA.
\item \textbf{GluonTS \citep{gluonts}} For this dataset, only Taxi, Uber TLC Daily, Uber TLC Hourly, Wiki-Rolling, and M5 are included. The rest of the datasets are already included in the Monash dataset.
\item \textbf{LargeST \citep{liu2023largest}} This dataset contains traffic  datasets from California Department of Transportation Performance Measurement System (PeMS). PeMS  includes PEMS03, PEMS04, PEMS07, PEMS08, PEMS Bay, and the well-known Traffic dataset.
\item \textbf{LibCity \citep{wang2023towards}} This is a collection of urban spatio-temporal datasets, while the spatial aspect is dropped.
\item \textbf{Monash} The Monash Time Series Forecasting Repository \citep{godahewa2021monash} is a comprehensive collection of diverse time series datasets. The test data for each dataset is the final horizon as the test set, while the forecast horizon is defined for each individual dataset. LOTSA includes the training data of Monash dataset, holding out the testing for in-distribution evaluation. Several datasets are included entirely in LOTSA: London Smart Meters, Wind Farms, Wind Power, Solar Power, Oikolab Weather, Covid Mobility, Extended Web Traffic, Kaggle Web Traffic Weekly, M1 Yearly, M1 Quarterly, M3 Yearly, M3 Quarterly, M4 Yearly, M4 Quarterly, Tourism Yearly. In our experiment evaluation, we do not fine-tune Delphyne on several datasets in Monash, due to that their training data is very short ($<20$ time steps) after splitting the data into test data and validation data.
\item \textbf{ProEnFo} ProEnFo \citep{wang2024benchmarks} is a dataset for load forecasting. Its data contains various covariates such as temperature, humidity, and wind speed.
\item \textbf{SubseasonalClimateUSA \citep{mouatadid2023subseasonalclimateusa}} This dataset offers climate time series data for subseasonal forecasting. LOTSA contains Subseasonal Precipitation, containing precipitation data from 1948 to 1978, and Subseasonal, which includes both precipitation and temperature data from 1979 to 2023.
\item 
\textbf{Other} LOTSA also contains datasets from miscellaneous sources, spanning from energy, econ/finance, sales and healthcare. Refer to Table 17 in \cite{woo2024unifiedtraininguniversaltime} for details. 
\end{enumerate}

\subsection{Financial Data}\label{ssec:fin_data}

By design, our data sampling samples time-series from the same dataset with the same sample ID; because of this, some of our datasets have the same time-series but are used in different contexts (sampled with different time series).
\begin{enumerate}
    \item \textbf{Single Currency Daily} This dataset includes 12 exchange rates, and each exchange rate is treated as a separate sample. For each exchange rate, we include the time series of the exchange rate and its returns, forward rates, and implied volatilities. We use 1W, 1M, 3M, 6M, 9M, 1Y, 18M, and 2Y for the tenors, and 0.1, 0.15, 0.25, 0.35, 0.5, 0.65, 0.75, 0.85, and 0.9 for the deltas. This dataset trains the model to properties across the spot, forward, and volatility surface.
    \item \textbf{Joint Currencies Daily} This data includes 68 currency pairs, the exchange rate and returns are the time series. This dataset is to allow our model to learn correlations across currencies; to this end, there is only one sample.
    \item \textbf{Currencies Monthly} This dataset includes 43 exchange rates, and each exchange rate is treated as a separate sample. We use the same columns as in Single Currency Daily except the returns are monthly returns.
    \item \textbf{Commodities Daily} This dataset includes 29 commodities, and each commodity is treated as a separate sample. Similar to exchange rates, we include the price and returns of the commodity, and the implied volatilities for 1M, 2M, 3M, 6M, 1Y, 18M, and 2Y for the tenors, and 90\%, 95.0\%, 97.5\%, 100.0\%, 102.5\%, 105.0\%, and 110.0\% for the moneyness.

\begin{table}[!bt]
  \centering
  \scalebox{0.95}{
  \begin{tabular}{ccc}
\toprule
BO1 & HO1 & QS1 \\
CC1 & JO1 & SB1 \\
CL1 & KC1 & SI1 \\
CO1 & KO1 & S \\
CT1 & LA1 & TZT1 \\
CU1 & LB1 & UXA1 \\
C & LP1 & W \\
FN1 & MO1 & XB1 \\
GC1 & NG1 & XW1 \\
HG1 & PL1 \\
\bottomrule
\end{tabular}
}
\caption{List of commodities tickers}
\label{tab:commodities}
\end{table}

    \item \textbf{Commodities Monthly} This dataset is identical to `Commodities Daily' except that the data is resampled to monthly level where we take monthly returns and the last value per month for the rest of the columns.
    \item \textbf{Joint Stock Returns} Similar to `Joint Currencies Daily', this dataset is to allow our model to learn correlations across stocks. This data includes the returns of 10,511 stocks. To ensure the correlations learned are more meaningful, we partitioned the stocks into 53 exchanges.
    \item \textbf{Daily ETFs Returns} Similar to `Joint Stock Returns', this dataset is to allow our model to learn correlations across ETFs. This data includes the returns of 28,837 ETFs. To ensure the correlations learned are more meaningful, we partitioned the stocks into 76 exchanges.
    \item \textbf{Company Data} Similar to `Single Currency Daily', this dataset is to allow our model to learn correlation across stock features. This data includes 10,511 stocks. For variates, we include stock returns, volume traded, quarterly company financials, consumer transaction data,  forward rates for the same tenors as 'Single Currency Daily', and implied volatilities for the same moneynesses as 'Commodities Daily' as well as 80\% and 120\% moneyness.
 \begin{table}[!bt]
  \centering
  \scalebox{0.95}{
  \begin{tabular}{ccc}
\toprule
is\_sg\&a\_expense & net\_income & is\_cogs\_to\_fe\_and\_pp\_and\_g \\
is\_sales\_and\_services\_revenues & is\_other\_operating\_expenses & is\_cog\_and\_services\_sold \\
is\_operating\_expn & ebit & sales\_rev\_turn \\
ebitda & short\_and\_long\_term\_debt & bs\_tot\_asset \\
bs\_cur\_liab & bs\_cur\_asset\_report & bs\_gross\_fix\_asset \\
total\_equity & bs\_pfd\_eqty\_\&\_hybrid\_cptl & bs\_inventories \\
bs\_cash\_near\_cash\_item & bs\_lt\_invest & bs\_net\_fix\_asset \\
bs\_acct\_note\_rcv & cash\_and\_marketable\_securities & enterprise\_value \\
net\_debt & sales\_to\_net\_fix\_asset & gross\_profit \\
num\_of\_employees & gross\_margin & historical\_market\_cap \\
avg\_age\_of\_assets\_in\_years & cf\_cap\_expend\_prpty\_add & cf\_cash\_from\_inv\_act \\
\bottomrule
\end{tabular}
}
\caption{List of company financials fields}
\label{tab:cofi_fields}
\end{table}

\begin{table}[!bt]
  \centering
  \scalebox{0.95}{
  \begin{tabular}{ccc}
\toprule
observed\_sales & observed\_transactions & observed\_unique\_customers \\
average\_transaction\_value & transactions\_per\_customer & sales\_per\_customer \\
\bottomrule
\end{tabular}
}
\caption{List of consumer transactions fields}
\label{tab:second_measure_fields}
\end{table}

    \item \textbf{Intraday Bars} 
    We include 15,817 global securities and for each five-minute interval (bar), we include the open, high, low, and closing price, the volume traded, and the number of trades. Since the securities have different open and close hours, to normalize the data, we drop specific five-minute intervals (a specify day of the week and time) for which the ticker has had zero trades through its life.
\end{enumerate}

\FloatBarrier
\section{Monash Time Series Forecast}
\subsection{Comparison Methods}
\textbf{Pre-trained Models.} For pre-trained models we report the zero-shot performance of MOIRAI \citep{woo2024unifiedtraininguniversaltime}. MOIRAI is a unified pre-trained foundation model for time-series analysis. Across the three versions of MOIRAI, we report it performance across MOIRAI$_\text{Base}$, which is roughly the same amount of parameters as our Delphyne models.

\textbf{Baselines.} Several traditional and statistical methods serve as the reported baseline for Monash, using the last observed value for the forecast. SES (Single Exponential Smoothing) applies a weighted average to past observations, with exponentially decreasing weights for older data points. Theta \citep{assimakopoulos2000theta} fits $\theta$-lines with exponential smoothing. 
Exponential Smoothing (ETS) is also a traditional statistical method.

\textbf{Non-deep Methods.} The non-deep learning methods include CatBoost (gradient boosting on decision trees) \citep{catboost}, (DHR)-ARIMA (dynamic harmonic regression), PR.

\textbf{Deep Methods.} Methods that including training neural networks include N-BEATS \citep{Oreshkin2020N-BEATS}, feed-forward neural network (FFMM), DeepAR \citep{salinas2020deepar}, N-BEATS \citep{Oreshkin2020N-BEATS}, WaveNet \citep{wavenet} and Transformer (Trans).

\textbf{GPT3.5 \& Llama2.} GPT3.5 and Llama2 are two versions of LLMTime. For GPT-3.5, we report the reproduced results by \cite{woo2024unifiedtraininguniversaltime}, as well as the original results by \cite{gruver2024LLMTime} run on Llama2.

\subsection{Full Comparison Results}

See Table \ref{tab:monash_results} for a comparison across baselines and Table \ref{tab:monash_delphyne} for comparison across different versions of Delphyne.
\begin{table}[htbp]
\centering
\caption{Full results of Monash Time Series Forecasting Benchmark. MAE is reported. The best result is in \textbf{bold}. "\textbf{Aggregated}" means that we take the
geometric mean of the MAE of each dataset divided by the MAE of the Naive approach (for zero-shot models only, fine-tuned performances are reported in Table \ref{tab:monash_delphyne}). }
\label{tab:monash_results}
\resizebox{1.0\textwidth}{!}{
\begin{tabular}%
{lccccccccccccccccccc}
\toprule
\textbf{Dataset} & \methoda-ZS & \methoda-FT & \textbf{MOIRAI} & \textbf{Naive} & \textbf{SES} & \textbf{Theta} & \textbf{TBATS} & \textbf{ETS} & \textbf{(DHR-)ARIMA} & \textbf{PR} & \textbf{CatBoost} & \textbf{FFNN} & \textbf{DeepAR} & \textbf{N-BEATS} & \textbf{WaveNet} & \textbf{Trans.} & \textbf{GPT3.5} & \textbf{LLaMA2}  \\ \midrule
M1 Monthly & 2153.37  & - & 2068.63 & 2707.75 & 2259.04 & 2166.18 & 2237.50 & 1905.28 & 2080.13 & 2088.25 & 2052.32 & 2162.58 & 1860.81 & \textbf{1820.37} & 2184.42 & 2723.88 & 2562.84 & - \\
M3 Monthly & 649.77 & - & 658.17 & 837.14 & 743.41 & \textbf{623.71} & 630.59 & 626.46 & 654.8 & 692.97 & 732 & 692.48 & 728.81 & 648.6 & 699.3 & 798.38 & 877.97 & - \\
M3 Other & 202.44 & - & 198.62 & 278.43 &  277.83 & 215.35 & \textbf{189.42} & 194.98 & 193.02 & 234.43 & 318.13 & 240.17 & 247.56 & 221.85 & 245.29 & 239.24 & 300.30 & - \\
M4 Monthly & 597.43 & \textbf{560.47} & 592.09 &  671.27 & 625.24 & 563.58 & 589.52 & 582.6 & 575.36 & 596.19 & 611.69 & 612.52 & 615.22 & 578.48 & 655.51 & 780.47 & 728.27 & - \\ 
M4 Weekly & 378.36  & 306.06 & 328.08 & 347.09 & 336.82 & 333.32 & 296.15 & 335.66 & 321.61 & 293.21 & 364.65 & 338.37 & 351.78 & \textbf{277.73} & 359.46 & 378.89 & 518.44 & - \\ 
M4 Daily & 223.54 & 181.89 & 192.66 & 180.83 & 178.27 & 178.86 & \textbf{176.6} & 193.26 & 179.67 & 181.92 & 231.36 & 177.91 & 299.79 & 190.44 & 189.47 & 201.08 & 266.52 & - \\ 
M4 Hourly & 218.85  & 211.93 & \textbf{209.87} & 1218.06 & 1218.06 & 1220.97 & 386.27 & 3358.10 & 1310.85 & 257.39 & 285.35 & 385.49 & 886.02 & 425.75 & 393.63 & 320.54 & 576.06 & - \\
Tourism Quarterly & 9487.13 & - & 17196.86 & 15845.10 & 15014.19 & \textbf{7656.49} & 9972.42 & 8925.52 & 10475.47 & 9902.58 & 10267.97 & 8881.04 & 9511.37 & 8640.56 & 9137.12 & 9521.67 & 16918.86 & 9311.98 \\
Tourism Monthly & 2615.26 & 2488.41 & 2862.06 & 5636.83 & 5302.10 & 4996.60 & 6940.08 & 5804.51 & 6022.21 & 5536.70 & 6071.62 & 5315.74 & \textbf{1871.69} & 2003.02 & 5998.22 & 4057.97 & 5608.81 & 3145.48 \\
CIF 2016(E+5) & \textbf{4.67} & - & 5.39 & 5.78 & 5.81 & 7.15 & 8.56 & 6.42 & 4.69 & 5.63 & 6.04 & 14.96 & 32.00 & 6.79 & 59.98 & 40.58 & 5.99 & 6.84 \\
 
Aus. Elec.  & 235.49 & 248.62 & \textbf{201.39} & 659.6 & 659.6 & 655.14 & 370.74 & 1282.99 & 1045.92 & 247.18 & 241.77 & 258.76  & 302.41 & 213.83 & 227.5 & 231.45 & 760.81 & 560.48  \\ 
Bitcoin(E+18) & 2.15 & 1.44 & 1.87 & 0.78 & 5.53 & 5.53 & 0.99 & 1.10 & 3.62 & \textbf{0.66} & 1.93 & 1.45 & 1.95 & 1.06 & 2.46 & 2.61  & 1.74 & 8.57 \\
Pedestrian Counts & 52.99 & 44.50 & 23.17 & 170.88 & 170.87 & 170.96 & 222.38 & 216.5 & 635.16 & 44.18 & 43.41 & 46.41 & 44.78 & 66.84 & 46.46 & 47.29 & 97.77 & 65.92 \\ 
Vehicle Trips & 17.68 & - & 21.85 & 31.42 & 29.98 & 30.76 & \textbf{21.21} & 30.95 & 30.97 & 27.24 & 22.61 & 22.93 & 22 & 28.16 & 24.15 & 28.01 & 31.48 & - \\ 
KDD cup & 30.87 & \textbf{29.96} & 39.09 & 42.13  & 42.04 & 42.06 & 39.2 & 44.88 & 52.2 & 36.85 & 34.82 & 37.16 & 48.98 & 49.1 & 37.08 & 44.46 & 42.72 & - \\
Weather & 2.05 & \textbf{1.79} & 1.8 & 2.36 & 2.24 & 2.51 & 2.3 & 2.35 & 2.45 & 8.17 & 2.51 & 2.09 & 2.02 & 2.34 & 2.29 & 2.03 & 2.17& 2.09 \\ 
NN5 Daily & \textbf{3.57} & 3.65 & 4.26 & 8.26 & 6.63 & 3.8 & 3.7& 3.72 & 4.41 & 5.47 & 4.22 & 4.06 & 3.94 & 4.92 & 3.97 & 4.17 & 7.10 & 6.67 \\ 
NN5 Weekly & 15.00 & \textbf{14.28} & 16.42 & 16.71 & 15.66 & 15.3 & 14.98 & 15.7 & 15.38 & 14.94 & 15.29 & 15.02 & 14.69& 14.19 & 19.34 & 20.34 & 15.76 & 15.60 \\
Carparts & 0.65 & - & 0.47 & 0.65 & 0.55 & 0.53 & 0.58& 0.56 & 0.56 & 0.41 & 0.53 & \textbf{0.39} & \textbf{0.39} & 0.98 & 0.4 & 0.39 & 0.44 & - \\
FRED-MD & 3806.16 & 2907.86 & 2679.29  & 2825.67 & 2798.22 & 3492.84 & 1989.97 & 2041.92 & 2957.11 & 8921.94 & 2475.68 & 2339.57 & 4264.36 & 2557.8 & 2508.4 & 4666.04 & 2804.64 & 
\textbf{1781.41} \\ 
Traffic Hourly & 0.02 & 0.02 & 0.02 & 0.03 & 0.03 & 0.03 & 0.04 & 0.03 & 0.04 & 0.02 & 0.02 & \textbf{0.01} & \textbf{0.01} & 0.02 & 0.02 & \textbf{0.01} & 0.03 & 0.02 \\ 
Traffic Weekly & 1.13 & 1.13 & 1.14 & 1.19 & \textbf{1.12} & 1.13 &1.17 & 1.14 & 1.22 & 1.13 & 1.17 & 1.15 & 1.18 & 1.11 & 1.2 & 1.42 & 1.15 & 1.15 \\ 
Rideshare & 1.12 & \textbf{1.11} & 1.39 & 6.29 & 6.29 & 7.62 & 6.45 & 6.29 & 3.37 & 6.3 & 6.07 & 6.59 & 6.28 & 5.55 & 2.75 & 6.29 & 6.28 & -  \\ 
Hospital & 19.08 & - & 19.4 & 24.07 & 21.76 & 18.54 & \textbf{17.43} & 17.97 & 19.6 & 19.24 & 19.17 & 22.86 &  18.25 & 20.18 & 19.35 & 36.19 & 25.68 & 22.75 \\ 
COVID Deaths & 174.35 & 137.76 & 126.11 & 353.71 & 353.71  & 321.32 & 96.29 & 85.59 & 87.77 & 347.98 & 475.15 & 144.14 & 201.98 & 158.81 & 1049.48 & 408.66 &  653.31 & \textbf{66.14}  \\ 
Temperature Rain & 6.27 & 5.14 & \textbf{5.08} & 9.39 & 8.18 & 8.22 & 7.14 & 8.21 & 7.19 & 6.13 & 6.76 & 5.56 & 5.37 & 7.28 & 5.81 & 5.24 & 6.37 & -\\ 
Sunspot & 3.51 & 0.41 & \textbf{0.08} & 3.93 & 4.93 & 4.93 & 2.57 & 4.93 & 2.57 & 3.83 & 2.27 & 7.97 & 0.77 & 14.47 & 0.17 & 0.13 & 5.07 & 0.28\\ 
Saugeen & 25.41  & \textbf{21.55} & 24.4 & 21.5 & 21.5 & 21.49 & 22.26 & 30.69 & 22.38 & 25.24 & 21.28 & 22.98 & 23.51 & 27.92 & 22.17 & 28.06 & 34.84 & 23.01 \\ 
US Births & 462.16 & 442.71 & 614.3 & 1152.67 & 1192.20 & 586.93 & \textbf{399} & 417.93 & 526.33 & 574.92 & 441.7 & 557.87 & 422 & 504.4 & 452.87 & 1374.99 & 638.82 \\ \hline
\textbf{Aggregated} & 0.68 & - & \textbf{0.58} & 1.0 & 1.03 & 0.96 & 0.78 & 0.91 & 0.93 & 0.81 & 0.78 & 0.77 & 0.76 & 0.79 & 0.77 & 0.82 & 1.01 & - \\ \hline
\end{tabular}
}
\end{table}

\begin{table}[htbp]
\centering
\caption{Delphyne model results of Monash Time Series Forecasting Benchmark. MAE is reported; for fine-tuning, the MAE is taken over 3 experimental runs and we report the mean $\pm$ std. The best result is in \textbf{bold}. "\textbf{Aggregated (Fine-tune)}" means that we take the
geometric mean of the MAE of each fine-tuned dataset divided by the MAE of the Naive approach. }
\label{tab:monash_delphyne}
\resizebox{1.0\textwidth}{!}{
\begin{tabular}{lcccccc}
\toprule
\textbf{Dataset} & \methodl-ZS & \methodf-ZS & \methoda-ZS  &  \methodl-FT & \methodf-FT  & \methoda-FT  \\ \toprule
M1 Monthly & 2252.507 & 2298.433 & \textbf{2153.370} & -&-&-\\ 
M3 Monthly & \textbf{643.874} & 817.461 & 649.765 & -&-&-\\
M3 Other & 209.257 & 493.340 & \textbf{202.444} & -&-&-\\
M4 Monthly & 620.923 & 697.144 & 597.434 & 569.709$\pm$7.662 & 563.145$\pm$9.496 & \textbf{560.472$\pm$2.022} \\ 
M4 Weekly & 340.878 & 379.246 & 378.363 & \textbf{296.282$\pm$4.909} & 319.959$\pm$38.095 & 306.055$\pm$5.195\\ 
M4 Daily & 210.110 & 201.094 & 223.536 & \textbf{174.349$\pm$0.727} & 182.327$\pm$3.155 & 181.884$\pm$5.742
\\ 
M4 Hourly & 234.718 & 1369.664 & 218.845 & 263.368$\pm$16.039 & 331.591$\pm$29.941 & \textbf{211.930$\pm$0.657}  \\
Tourism Quarterly & \textbf{9268.650} & 17143.935 & 9487.128 & -&-&- \\
Tourism Monthly & 2458.334 & 6074.311 & 2615.256 & \textbf{2360.693$\pm$123.850} & 2460.906$\pm$277.598 & 2488.407$\pm$205.380 \\ 
CIF 2016 (E+6) & 5.408 & \textbf{2.826} & 4.670 & -&-&- \\ 
Aus. Elec.  & 203.034 & 1795.900 & 235.490 & \textbf{199.854$\pm$0.200} & 235.263$\pm$5.601 & 248.615$\pm$0.959 \\ 
Bitcoin (E+18) & 1.928 & 1.871  & 2.152 & 1.255$\pm$0.174 & \textbf{1.185$\pm$0.861} & 1.437$\pm$0.332\\ 
Pedestrian Counts & 44.367 & 170.285 & 52.987 & \textbf{41.917$\pm$0.532} & 58.494$\pm$5.365 & 44.505$\pm$1.217 \\ 
Vehicle Trips & 18.327 & 20.227 & \textbf{17.682} & -&-&- \\ 
KDD cup & 30.045 & 35.768 & 30.868 & 30.029$\pm$0.019 & \textbf{28.822$\pm$0.110} & 29.964$\pm$0.039 \\ 
Weather &2.053 & 2.495 & 2.052& \textbf{1.776$\pm$0.014} & 1.807$\pm$0.036 & 1.792$\pm$0.010\\ 
NN5 Daily & 3.596 & 3.622 & 3.574 & 3.566$\pm$0.044 & 
\textbf{3.474$\pm$0.017} & 3.648$\pm$0.063\\ 
NN5 Weekly & 14.899 & 15.945 & 14.999 & \textbf{14.109$\pm$0.045} & 14.189$\pm$0.047 & 14.280$\pm$0.087\\ 
Carparts & 0.656 & 0.662 & \textbf{0.652} & -&-&- \\ 
FRED-MD & 2868.493 & 3510.298 & 3806.159 & \textbf{2720.094$\pm$101.289} & 4008.423$\pm$198.970 & 2907.856$\pm$200.807 \\ 
Traffic Hourly & 0.018 & 0.038 & 0.018 & \textbf{0.015$\pm$0.000} & 0.016$\pm$0.000 & 0.017$\pm$0.001 \\ 
Traffic Weekly & 1.121 & 1.163 & 1.125  & \textbf{1.108$\pm$0.007} & 1.123$\pm$0.005 & 1.131$\pm$0.010\\ 
Rideshare & 1.256 & 1.676 & 1.123 & 1.221$\pm$0.010 & \textbf{1.110$\pm$0.001} & 1.111$\pm$0.001\\ 
Hospital & \textbf{19.029} & 22.427 & 19.081 & -&-&-\\ 
COVID Deaths & 154.175 & 385.451 & 174.354 & \textbf{135.791$\pm$21.305} & 180.857$\pm$23.128 & 137.716$\pm$4.510 \\ 
Temperature Rain & 6.794 & 7.515 & 6.267 & 5.341$\pm$0.160 & 5.482$\pm$0.112 & \textbf{5.142$\pm$0.006} \\ 
Sunspot &  \textbf{3.163} &  9.458 & 3.507 & 0.328$\pm$0.035 & 0.455$\pm$0.072 & 0.410$\pm$0.015\\ 
Saugeen & 24.281 & 23.757 & 25.410 & 24.780$\pm$0.088 & 23.258$\pm$0.209 & \textbf{21.552$\pm$0.169} \\ 
US Births &443.349 & 462.390 & 463.157 & \textbf{365.175$\pm$42.777} & 425.495$\pm$9.022 & 442.705$\pm$0.801\\ \hline
\textbf{Aggregated} \\
\textbf{Fine-tune} & 0.623 & 0.929 & 0.626 & \textbf{0.514} &0.562 & 0.536 \\ \hline
\end{tabular}
}
\end{table}

\FloatBarrier
\section{Financial Tasks}

The hyperparameters we tuned are in Table \ref{tbl:stock_hparams}. Note, for MOMENT \citep{goswami2024moment}, we use  patch length of 8 and the context length to 512 since these are fixed by the model. Similarly, for TTM \citep{ekambaram2024tinytimemixersttms}, we used a context length of 512 since this is fixed by the model; we also used a head dropout of 0.7 (as suggested in the paper).

\begin{table}[htbp]
  \centering
  \caption{Hyperparameter search values for financial tasks.}
  \resizebox{0.5\columnwidth}{!}{
    \begin{tabular}{lcc}
    \toprule
          & \textbf{Hyperparameter} & \textbf{Values} \\
    \midrule
    Delphyne & learning rate & \(\{1e-5,5e-4,1e-4\}\) \\
          & dropout & \(\{0.1,0.2,0.3\}\) \\
    \midrule
    MOIRAI  & learning rate & \(\{1e-5,5e-4,1e-4\}\) \\
        & patch size & \(\{16, 32\}\) \\
    \midrule
    TTM & learning rate & \(\{1e-5,5e-4,1e-4, 1e-3\}\) \\
    \midrule
    PatchTST & patch size & \(\{1,4, 16, 32\}\) \\
          & hidden size & \(\{64,128,256\}\) \\
          & dropout & \(\{0.0,0.1,0.2\}\) \\
    \bottomrule
    \end{tabular}%
  }
  \label{tbl:stock_hparams}%
\end{table}%

\subsection{Calibration Analysis for Stocks Task}\label{ssec:cov_stocks}

In addition to evaluating the models with NLL, we can also evaluate the coverage statistics: \textbf{given a forecasted quantile $q$, what percentage of the observations are less than that value?} In Table \ref{tab:stocks_nll_calib_results}, we present the results. We see that \methoda-FT performs the best or second-best in nearly all of the quantiles.

\begin{table}[htbp]
\centering
\caption{Results of stock risk analysis for zero-shot versus fine-tuning with stocks coverage statistic. We \textbf{bold} the results that are closest in absolute error to the optimal coverage.}
\vspace{-0.1in}
\label{tab:stocks_nll_calib_results}
\begin{tabular}{lccccc}
\hline
\textbf{Model}      & \textbf{Q10} & \textbf{Q25} & \textbf{Q50} & \textbf{Q75}& \textbf{Q90}  \\
\hline
\textbf{Optimal}   & 0.100 & 0.250 & 0.500 & 0.750 & 0.900       \\
\hline
\textbf{\methoda}-ZS & 0.109 & 0.239 & \underline{0.502} & \underline{0.749} & 0.892  \\
\textbf{\methodf}-ZS & \underline{0.104} & \textbf{0.252} & 0.515 & 0.777 & 0.907  \\
\textbf{\methodl}-ZS & \textbf{0.097} & 0.238 & 0.525 & 0.790 & 0.906  \\
\hline
\textbf{\methoda}-FT & 0.106 & \underline{0.246} & \textbf{0.501} & \textbf{0.750} & \textbf{0.900}  \\
\textbf{\methodf}-FT & 0.086 & 0.216 & 0.487 & 0.768 & \underline{0.903}  \\
\textbf{\methodl}-FT & 0.089 & 0.216 & 0.510 & 0.794 & 0.916  \\
\hline
\textbf{MOIRAI}-ZS      & 0.108 & 0.215 & 0.445 & 0.732 & 0.877       \\
\textbf{MOIRAI}-FT      & 0.083 & 0.205 & 0.493 & 0.794 & 0.917       \\
\textbf{GARCH} & 0.111 & 0.269 & 0.561 & 0.790 & 0.913 \\
\textbf{PatchTST} &  0.114& 0.264& 0.509& 0.748& 0.896       \\
\hline
\end{tabular}
\end{table}

\begin{table}[htbp]
\centering
\caption{Full results for zero-shot versus fine-tuning for predicting next-day stock squared-returns (variance) data. MSE is reported; for fine-tuning, the MSE is taken over 3 experimental runs and we report the mean $\pm$ std.}
\vspace{-0.1in}
\label{tab:stock_var_results}
\begin{tabular}{lcc}
\hline
\textbf{Model}      & \textbf{MSE ZS} & \textbf{MSE FT} \\
\hline
\textbf{\methoda} & 37.792          & 37.810$\pm$0.105  \\
\textbf{\methodf} & \underline{37.653}          & 38.616$\pm$1.566  \\
\textbf{\methodl} & \textbf{37.591}          & 38.246$\pm$0.598  \\
\textbf{MOIRAI}      & 41.428          & 40.502$\pm$0.046          \\
\textbf{MOMENT}      & 46.006          & 37.935$\pm$0.179          \\
\textbf{TTM}         & 44.918          & 44.360$\pm$0.004          \\
\textbf{PatchTST}    & -          & 51.705$\pm$11.467          \\
\textbf{GARCH}       & 41.517          & -               \\
\hline
\end{tabular}
\vspace{-0.2in}
\end{table}

\begin{table}[htbp]
\centering
\caption{Full results for zero-shot versus fine-tuning for next-day stock-returns risk analysis. NLL is reported; for fine-tuning, the NLL is taken over 3 experimental runs and we report the mean $\pm$ std.}
\vspace{-0.1in}
\label{tab:stocks_nll_results}
\begin{tabular}{lcc}
\hline
\textbf{Model}      & \textbf{NLL ZS} & \textbf{NLL FT} \\
\hline
\textbf{\methoda} & 1.762          & \textbf{1.741$\pm$0.002}  \\
\textbf{\methodf} & 1.750          & \underline{1.746$\pm$0.001}  \\
\textbf{\methodl} & 1.775          & 1.757$\pm$0.005  \\
\textbf{MOIRAI}      & 1.776          & 1.788$\pm$0.001          \\
\textbf{GARCH}       & 1.752          & -               \\
\textbf{PatchTST}  & -  &  1.751$\pm$0.005\\
\hline
\end{tabular}
\end{table}

\begin{table}[htbp]
\centering
\caption{Full results for zero-shot versus fine-tuning for predicting bars log-volume data (longer horizon, 78 timesteps prediction for 5-minute intervals). MSE is reported; for fine-tuning, the MSE is taken over 3 experimental runs and we report the mean$\pm$std.}
\vspace{-0.1in}
\label{tab:results}
\begin{tabular}{lcc}
\hline
\textbf{Model}      & \textbf{MSE ZS} & \textbf{MSE FT} \\
\hline
\textbf{\methoda} & 0.728           & 0.551$\pm$0.017     \\
\textbf{\methodf} & 0.965           & \textbf{0.530$\pm$0.01}      \\
\textbf{\methodl} & 0.930           & 0.557$\pm$0.002     \\
\textbf{MOIRAI}      & 0.765           & 0.621$\pm$0.003             \\
\textbf{MOMENT}      & 0.775           & 0.838$\pm$0.028             \\
\textbf{TTM}         & 0.714           & 0.600$\pm$0.001             \\
\textbf{PatchTST}    &    -           & \underline{0.534}             \\
\textbf{Last Value}  & 0.602           & -                 \\
\hline
\end{tabular}
\end{table}

\begin{table}[htbp]
\centering
\caption{Nowcasting results for zero-shot vs. fine-tuning for company sales growth data. MAE is reported; for fine-tuning, the MAE is taken over 3 experimental runs and we report the mean $\pm$ std.}
\vspace{-0.1in}
\label{tab:bsm_results}
\begin{tabular}{lcc}
    \toprule
    \textbf{Model}      & \textbf{MAE ZS} & \textbf{MAE FT} \\
    \midrule
    \textbf{\methoda}    & 0.099         & \textbf{0.071$\pm$0.002} \\
    \textbf{\methodf}    & 0.128       & 0.079$\pm$0.003  \\
    \textbf{\methodl}    & 0.101        & \underline{0.073$\pm$0.001}  \\
    \textbf{MOIRAI}      & 0.091    & 0.093$\pm$0.001     \\
    \textbf{Baseline}      & 0.100    & -     \\
    \bottomrule
    \end{tabular}
\end{table}

\FloatBarrier
\section{Long-term Forecasting Experiment}
\label{sec:long-term-forecasting-experiment-details}

For fine-tuning Delphyne, we use a learning rate of 5e-5, dropout of 0.2, batch size of 64, and a linear warmup for the learning rate of 50 steps. For all datasets, we use a context length of 1000. We use early-stopping based on the validation loss. Due to Electricity and Weather being large datasets, we randomly sample $32 \times 500$ rows from the validation set for early-stopping. For long-term forecasting experiment, we do not conduct any additional hyperparameter searching, although that could lead to improved performances.

\subsection{Comparison Methods}

\textbf{Zero-Shot Methods.} For zero-shot methods, we report TTM$_\text{A}$, the best and largest model presented in \cite{ekambaram2024tinytimemixersttms}. Since the authors have only published models with a forecast length of 96, we are limited to reporting the MSE based on their reported results. For MOIRAI \citep{woo2024unifiedtraininguniversaltime}, we again report the performance of MOIRAI$_\text{Base}$, which has similar number of parameters as our model. For TimesFM \citep{das2024timesfm}, we follow their demonstration \footnote{We use the following script and set different forecast lengths in \url{https://github.com/google-research/timesfm/blob/master/notebooks/finetuning.ipynb}.} and report the MSE and MAE results for their checkpoint "google/timesfm-1.0-200m" in Huggingface. 

\textbf{Linear-Probing.} We directly report the linear probing results from MOMENT's experiments \citep{goswami2024moment}, which include baseline results for both Time-LLM \citep{jin2024timellm} and GPT4TS \citep{zhou2023ofa}.

\textbf{Full-Shot.} The full-shot results are obtained from \cite{goswami2024moment}. Within the full-shot results, PatchTST \citep{nie2023patchtst}, DLinear \citep{elfwing2017sigmoidweightedlinearunitsneural}, TimesNet \citep{wu2023timesnet}, FEDFormer \citep{zhou2022fedformerfrequencyenhanceddecomposed}, Stationary \citep{zhou2023ofa}, LightTS \citep{lightts} and N-BEATS \citep{Oreshkin2020N-BEATS} are reported.

\subsection{Full Comparison Results}
Table \ref{tab:ett_overall} shows an comparison across different models and Table \ref{tab:ett_zs_ft} shows the comparison across different versions of Delphyne.

\begin{table}[htbp]
\centering
\caption{Zero-shot and Full-shot Results for Delphyne and Other Models}
\resizebox{\textwidth}{!}{
\begin{tabular}{lccccccccccccccccccccccccccccccccc}
\toprule
\multirow{4}{*}{\textbf{Dataset}} & \multirow{4}{*}{\textbf{Horizon}} & \multicolumn{8}{c}{\textbf{Zero-shot}} &
\multicolumn{6}{c}{\textbf{Linear-Probing}} & \multicolumn{2}{c}{\textbf{Fine-tuned}}  & \multicolumn{12}{c}{\textbf{Full-shot}}\\
\cmidrule(lr){3-10} \cmidrule(lr){11-16} \cmidrule(lr){17-18}
\cmidrule(lr){19-32}
& & \multicolumn{2}{c}{\textbf{TTM}} & \multicolumn{2}{c}{MOIRAI} & \multicolumn{2}{c}{\textbf{TimesFM}
} &\multicolumn{2}{c}{\methoda-ZS} & 
\multicolumn{2}{c}{\textbf{MOMENT}} &  \multicolumn{2}{c}{\textbf{Time-LLM}} &
\multicolumn{2}{c}{\textbf{GPT4TS}} &
\multicolumn{2}{c}{\methoda-FT} & 
\multicolumn{2}{c}{\textbf{PatchTST}} & \multicolumn{2}{c}{\textbf{DLinear}} & \multicolumn{2}{c}{\textbf{TimesNet}} & \multicolumn{2}{c}{\textbf{FEDFormer}} & 
\multicolumn{2}{c}{\textbf{Stationary}} & 
\multicolumn{2}{c}{\textbf{LightTS}} & \multicolumn{2}{c}{\textbf{N-BEATS}}\\
 & & \textbf{MSE} & \textbf{MAE} & \textbf{MSE} & \textbf{MAE} &  \textbf{MSE} & \textbf{MAE} & \textbf{MSE} & \textbf{MAE} & \textbf{MSE} & \textbf{MAE} & \textbf{MSE} & \textbf{MAE} & \textbf{MSE} & \textbf{MAE} & \textbf{MSE} & \textbf{MAE}
 & \textbf{MSE} & \textbf{MAE} & \textbf{MSE} & \textbf{MAE} & \textbf{MSE} & \textbf{MAE} & \textbf{MSE} & \textbf{MAE} & \textbf{MSE} & \textbf{MAE} & \textbf{MSE} & \textbf{MAE} & \textbf{MSE} & \textbf{MAE}\\
\midrule
\multirow{4}{*}{ETTh1} & 96  & 0.359 & -  & 0.384 & 0.402 & 0.405 & 0.432 & 0.398 & 0.417 & 0.387 & 0.410  & 0.408 & 0.429 & 0.376 & 0.397 & 0.376 & 0.390 &  0.370 & 0.399 & 0.375 & 0.399 & 0.384 & 0.402 & 0.376 & 0.419 & 0.513 & 0.491 & 0.424 & 0.432 & 0.399 & 0.428 \\
& 192 & 0.389 & - &  0.425 & 0.429 & 0.459 & 0.432 & 0.440 & 0.444 & 0.410 & 0.426 & - & - & 0.416 & 0.418 & 0.432 &  0.440 & 0.413 & 0.421 & 0.405 & 0.416 & 0.436 & 0.429 & 0.420 & 0.448 & 0.534 & 0.504 & 0.475 & 0.462 & 0.451 & 0.464 \\
& 336 & 0.405 & - &  0.456 & 0.450  & 0.534 & 0.458  & 0.474 & 0.464 & 0.422 & 0.437 & - & - & 0.442& 0.433 & 0.452 & 0.481 & 0.422 & 0.436 & 0.439 & 0.443 & 0.491 & 0.469 & 0.459 & 0.465 & 0.588 & 0.535 & 0.518 & 0.488 & 0.498 & 0.500 \\
& 720 & 0.448 & -  & 0.470 & 0.473 & 0.513 & 0.482 & 0.482 & 0.479 & 0.454 & 0.472&  0.523 & 0.514 & 0.477 & 0.456 & 0.498 & 0.454  & 0.447 & 0.466 & 0.472 & 0.490 & 0.521 & 0.500 & 0.506 & 0.507 & 0.643 & 0.616 & 0.547 & 0.533 & 0.608 & 0.573\\
& \textbf{Avg} & 0.402 & - & 0.434 & 0.439 & 0.476 & 0.451 & 0.449 & 0.450 & 0.418 & 0.436 & - & - & 0.428 & 0.426 & 0.440 & 0.441 & 0.413 & 0.431 & 0.423 & 0.437 & 0.457 & 0.449 & 0.440 & 0.460 & 0.570 & 0.537 & 0.491 & 0.479 & 0.489 & 0.491  \\
\midrule
\multirow{4}{*}{ETTh2} & 96  & 0.264 & -& 0.277 & 0.327 & 0.311 & 0.344 & 0.303 & 0.352 & 0.288 & 0.345 & 0.285 & 0.348& 0.285 & 0.342 & 0.281 & 0.300 & 0.274 & 0.336 & 0.289 & 0.353 & 0.340 & 0.374 & 0.358 & 0.397 & 0.476 & 0.458 & 0.397 & 0.437 & 0.327 & 0.387\\
& 192 & 0.321 & - & 0.340 & 0.374 & 0.395 & 0.393 & 0.371 & 0.397 & 0.349 & 0.386  & - & - & 0.354 & 0.389  & 0.342 & 0.351 & 0.339 & 0.379 & 0.383 & 0.418 & 0.402 & 0.414 & 0.429 & 0.439 & 0.512 & 0.493 & 0.520 & 0.504 & 0.400 & 0.435\\
& 336 & 0.351 & - & 0.371 & 0.402 & 0.384 & 0.402 & 0.400 & 0.420 & 0.369 & 0.408  & - & - & 0.373 &  0.407  & 0.382 & 0.393 & 0.329 & 0.380 & 0.448 & 0.465 & 0.452 & 0.452 & 0.496 & 0.487 & 0.552 & 0.551 & 0.626 & 0.559 & 0.747 & 0.599\\
& 720 & 0.395 & - & 0.394 & 0.426 & 0.526 & 0.484 & 0.428 & 0.446 & 0.403 & 0.439 & - & - & 0.406 & 0.441 & 0.401 & 0.381 &0.379 & 0.422 & 0.605 & 0.551 & 0.462 & 0.468 & 0.463 & 0.474 & 0.562 & 0.560 & 0.863 & 0.672 & 1.454 & 0.847 \\
& \textbf{Avg} &  0.327 & - & 0.346 & 0.382 & 0.404 & 0.406 & 0.375 & 0.404 & 0.352 & 0.395 & - & - & 0.355 & 0.395 & 0.352 & 0.356 & 0.330 & 0.379 & 0.431 & 0.447 & 0.414 & 0.427 & 0.437 & 0.449 & 0.526 & 0.516 & 0.602 & 0.543 & 0.732 & 0.568\\
\midrule
\multirow{4}{*}{ETTm1} & 96  & 0.318 & - &  0.335  & 0.360 & 0.332 & 0.351 & 0.439 & 0.399 & 0.293 &0.349 &  0.384 & 0.403 & 0.292 & 0.346 & 0.318 & 0.320 & 0.290 &  0.342  & 0.299 & 0.343& 0.338 &0.375& 0.379 &0.419 &0.386& 0.398 &0.374& 0.400& 0.318& 0.367 \\
& 192 & 0.354 & - & 0.366 & 0.379 & 0.387 & 0.389 & 0.483 & 0.425 & 0.326 &  0.368  & - & - & 0.332 & 0.372 & 0.339 & 0.354 & 0.332& 0.369& 0.335& 0.365& 0.374 &0.387& 0.426& 0.441 &0.459 &0.444 &0.400& 0.407& 0.355& 0.391\\
& 336 & 0.376 & - & 0.391 & 0.394 & 0.427 & 0.420 & 0.512 & 0.445 & 0.352 & 0.384 & - & - & 0.366 & 0.394 & 0.377 & 0.384 & 0.366& 0.392& 0.369& 0.386& 0.410 &0.411& 0.445& 0.459 &0.495 &0.464& 0.438& 0.438 &0.401 &0.419 \\
& 720 & 0.398 & - & 0.434 & 0.419 &0.532 & 0.472 & 0.572 & 0.445 & 0.445 & 0.462  & 0.437 & 0.429 & 0.417 & 0.421 & 0.421 & 0.402 & 0.416 &0.420& 0.425& 0.421& 0.478& 0.450& 0.543 &0.490& 0.585& 0.516& 0.527 &0.502& 0.448& 0.448\\
& \textbf{Avg} & 0.338 & - & 0.382 & 0.388 & 0.420 & 0.408 & 0.501 & 0.429 & 0.354 & 0.391 & - & - & 0.352 & 0.383 & 0.364 & 0.365 & 0.351 & 0.381 & 0.357 & 0.379 & 0.400 & 0.406 & 0.448 & 0.452 & 0.481 & 0.456 & 0.435 & 0.437 & 0.381 & 0.406 \\
\midrule
\multirow{4}{*}{ETTm2} & 96  & 0.169 & - & 0.195 & 0.269 & 0.189 & 0.257 & 0.211 & 0.294 & 0.243 & 0.170  & 0.181 & 0.269 & 0.173& 0.262 & 0.159 & 0.179 & 0.165 &0.255& 0.167& 0.269& 0.187& 0.267& 0.203 &0.287 &0.192&0.274 &0.209& 0.308 &0.197& 0.271\\
& 192 & 0.223 & - & 0.247 & 0.303 & 0.284 & 0.314 & 0.278 & 0.338 & 0.279 & 0.285  & - & - &0.229& 0.301 & 0.216 & 0.233 & 0.220& 0.292& 0.224& 0.303& 0.249 &0.309& 0.269& 0.328 &0.280 &0.339 &0.311& 0.382& 0.285& 0.328\\
& 336 & 0.276 & - & 0.291 & 0.333 & 0.467 & 0.396 & 0.343 & 0.377 & 0.227 & 0.297  & - & - & 0.286 & 0.341 & 0.271 & 0.298 & 0.274 &0.329& 0.281 &0.342& 0.321& 0.351& 0.325 &0.366 &0.334& 0.361& 0.442& 0.466& 0.338& 0.366\\
& 720 & 0.342 & - & 0.355 & 0.377 & 0.460 & 0.445 & 0.459 & 0.441 & 0.275 & 0.328  & 0.366 & 0.388 & 0.378 & 0.401 & 0.352 & 0.317 & 0.362 &0.385 &0.397 &0.421& 0.408& 0.403 &0.421 &0.415& 0.417& 0.413& 0.675 &0.587& 0.395& 0.419 \\
& \textbf{Avg} & 0.264 & - & 0.272 & 0.321 & 0.350 & 0.353 & 0.323 & 0.363 & 0.256 & 0.270 & - & - & 0.266 & 0.326 & 0.250 & 0.257 & 0.255 & 0.315 & 0.267 & 0.334 & 0.291 & 0.333 & 0.305 & 0.350 & 0.306 & 0.347 & 0.409 & 0.436 & 0.304 & 0.347 \\
\midrule
\multirow{4}{*}{Electricity} & 96  & 0.152 & - & 0.158 & 0.248 & 0.116 & 0.210 & 0.164 & 0.260 & 0.136 & 0.233  & - &-& 0.139 &0.238 & 0.143 & 0.176   &0.129 &0.222 &0.140& 0.237& 0.168& 0.272& 0.193 &0.308& 0.169 &0.273& 0.207 &0.307 &0.131 &0.228 \\
& 192 & 0.179 & - & 0.174 & 0.263 & 0.138 & 0.229 & 0.181 & 0.276 & 0.152 & 0.247  & - & - & 0.153 & 0.251 & 0.159 & 0.192 &  0.157 &0.240& 0.153 &0.249& 0.184& 0.289& 0.201& 0.315 &0.182& 0.286 &0.213 &0.316& 0.153& 0.248 \\
& 336 & 0.193 & - & 0.191 & 0.278 & 0.157 & 0.251 & 0.202 & 0.296 &  0.167 & 0.264 & - & - & 0.169 & 0.266 & 0.173 & 0.207 &0.163 & 0.259&  0.169& 0.267 &0.198& 0.300 &0.214 &0.329& 0.200& 0.304 &0.230 &0.333& 0.170& 0.267 \\
& 720 & 0.243 & - & 0.229 & 0.307 & 0.211 & 0.292 & 0.260 & 0.340 & 0.205 & 0.295  & - & - & 0.206 & 0.297 & 0.206 & 0.238 & 0.197& 0.290 &0.203& 0.301 &0.220& 0.320& 0.246& 0.355& 0.222& 0.321 &0.265& 0.360& 0.208& 0.298\\
& \textbf{Avg} & 0.160 & - & 0.188 & 0.274 & 0.156 & 0.246 & 0.202 & 0.293 & 0.165 & 0.260 & - & - & 0.167 & 0.263 & 0.170 & 0.203 & 0.162 & 0.253 & 0.166 & 0.264 & 0.193 & 0.295 & 0.214 & 0.327 & 0.193 & 0.296 & 0.229 & 0.329 & 0.166 & 0.260 \\
\midrule
\multirow{4}{*}{Weather} & 96  & 0.159 & - & 0.167 & 0.203 & 0.117 & 0.156 & 0.188 & 0.248 & 0.154 &0.209  & - & - & 0.162 & 0.212 & 0.140 & 0.157 &  0.149& 0.198 &0.176& 0.237 &0.172& 0.220 &0.217 &0.296 &0.173& 0.223& 0.182& 0.242& 0.152 &0.210\\
& 192 & 0.203 & - & 0.197 & 0.248 & 0.165 & 0.205 & 0.265 & 0.309 & 0.209 & 0.214  & -  & -&  0.204 & 0.248 & 0.187 & 0.204 & 0.194 &0.241& 0.220 &0.282& 0.219& 0.261& 0.276& 0.336 &0.245 &0.285& 0.227 &0.287& 0.199& 0.260\\
& 336 & 0.247 & - & 0.256 & 0.276 & 0.256 & 0.276 & 0.409 & 0.374 &  0.246 & 0.285  & - & - & 0.254 & 0.286 & 0.242 & 0.255 & 0.245 &0.282& 0.265& 0.319& 0.280 &0.306& 0.339& 0.380 &0.321& 0.338& 0.282& 0.334& 0.258 & 0.311\\
& 720 & 0.314 & - & 0.321 & 0.323 & 0.388 & 0.392 & 0.612 & 0.460 & 0.315 & 0.336  & - & - & 0.326& 0.337 & 0.316 & 0.323 & 0.314& 0.334 &0.333 &0.362 &0.365 &0.359 &0.403& 0.428 &0.414 &0.410& 0.352& 0.386 &0.331 &0.359\\
& \textbf{Avg} &0.233 & - & 0.235 & 0.263 & 0.232 & 0.257 & 0.369 & 0.348 & 0.230 & 0.261 & - & - & 0.237 & 0.271 & 0.221 & 0.235 & 0.226 & 0.264 & 0.249 & 0.300 & 0.259 & 0.286 & 0.309 & 0.360 & 0.288 & 0.314 & 0.261 & 0.312 & 0.235 & 0.286\\
\bottomrule
\end{tabular}}
\label{tab:ett_overall}
\end{table}

\begin{table}[htbp]
\centering
\caption{Full results of long sequence forecasting experiments for zero-shot versus fine-tuning. \methodf underperforms both model due to lack of similar dataset in pre-training. Both \methoda and \methodl perform comparatively well after fine-tuning.}
\vspace{-0.1in}
\label{tab:results}
\scalebox{0.57}{
\begin{tabular}{cccccccccccccc}
\toprule
\multirow{2}{*}{\textbf{Dataset}} & & \multicolumn{2}{c}{\textbf{\methodl-ZS}} & \multicolumn{2}{c}{\textbf{\methodf-ZS}}  &\multicolumn{2}{c}{\textbf{\methoda-ZS}} & \multicolumn{2}{c}{\textbf{\methodl-FT}} & \multicolumn{2}{c}{\textbf{\methodf-FT}} & \multicolumn{2}{c}{\textbf{\methoda-FT}}\\
\cmidrule(lr){3-4} \cmidrule(lr){5-6} \cmidrule(lr){7-8} \cmidrule(lr){9-10}
\cmidrule(lr){11-12}
\cmidrule(lr){13-14}
 & & \textbf{MSE} & \textbf{MAE} & \textbf{MSE} & \textbf{MAE} & \textbf{MSE} & \textbf{MAE} &  \textbf{MSE} & \textbf{MAE} & \textbf{MSE} & \textbf{MAE} & \textbf{MSE} & \textbf{MAE}  \\
\midrule
\multirow{ 4}{*}{\textbf{ETTh1}} & 96 &  0.388 & 0.418 & 1.179 & 0.690 & 0.398 & 0.417 & 0.384$\pm$0.001 & 0.400$\pm$0.023 & 0.403$\pm$0.003 & 0.409$\pm$0.011 & \textbf{0.376$\pm$0.001} & \textbf{0.390$\pm$0.020} \\
 & 192 & 0.437 & 0.452 & 1.405 & 0.769 & 0.440 & 0.444 & 0.445$\pm$0.009 & 0.457$\pm$0.010 & 0.456$\pm$0.023 & 0.493$\pm$0.032 & \textbf{0.432$\pm$0.011} & \textbf{0.440$\pm$0.006} \\
 & 336 & 0.502 & 0.483 & 1.677 & 0.857 & 0.474 & \textbf{0.464} & 0.470$\pm$0.001 & 0.490$\pm$0.029 & 0.487$\pm$0.040 & 0.484$\pm$0.036 & \textbf{0.452$\pm$0.006} & 0.481$\pm$0.035\\
 & 720 & 0.598 & 0.526 & 2.203 & 1.025 & \textbf{0.482} & 0.479 & 0.534$\pm$0.013 & 0.489$\pm$0.067 & 0.517$\pm$0.013 & 0.484$\pm$0.042 & 0.498$\pm$0.023 & \textbf{0.454$\pm$0.039} \\
 \midrule
 
\multirow{ 4}{*}{\textbf{ETTh2}} & 96 & 0.297 & 0.354 & 0.440 & 0.433 & 0.303 & 0.352 & 0.283$\pm$0.005 & 0.302$\pm$0.032 & 0.328$\pm$0.013 & 0.351$\pm$0.045 & \textbf{0.281$\pm$0.006} & \textbf{0.300$\pm$0.031} \\
 & 192 & 0.368 & 0.405 & 0.612 & 0.513 & 0.371 & 0.397 & 0.352$\pm$0.007 & 0.364$\pm$0.014 & 0.402$\pm$0.014 & 0.393$\pm$0.012 & \textbf{0.342$\pm$0.001} & \textbf{0.351$\pm$0.014} \\
 & 336 & 0.412 & 0.434 & 0.799 & 0.592 & 0.400 & 0.420 & 0.383$\pm$0.002 & \textbf{0.389$\pm$0.008} & 0.406$\pm$0.017 & 0.420$\pm$0.012 & \textbf{0.382$\pm$0.009} & 0.393$\pm$0.007\\
 & 720 & 0.448 & 0.462 & 1.075 & 0.689 & 0.428 & 0.446 & 0.422$\pm$0.015 & 0.403$\pm$0.043 & 0.433$\pm$0.011 & 0.417$\pm$0.028 & \textbf{0.401$\pm$0.004} & \textbf{0.381$\pm$0.028} \\
 \midrule
\multirow{ 4}{*}{\textbf{ETTm1}} & 96 & 0.599 & 0.455 & 1.498 & 0.711 & 0.439 & 0.399 & \textbf{0.295$\pm$0.004} & \textbf{0.314$\pm$0.028} & 0.313$\pm$0.001 & 0.324$\pm$0.014 & 0.318$\pm$0.009 & 0.320$\pm$0.013 \\
 & 192 & 0.715 & 0.502 & 1.730 & 0.780 & 0.483 & 0.425 & 0.348$\pm$0.006 & \textbf{0.350$\pm$0.007} & 0.342$\pm$0.005 & 0.350$\pm$0.014 & \textbf{0.339$\pm$0.005} & 0.354$\pm$0.021  \\
 & 336 & 0.795 & 0.534 & 2.006 & 0.861 & 0.512 & 0.445 & \textbf{0.363$\pm$0.004} & \textbf{0.382$\pm$0.024} & 0.385$\pm$0.014 & 0.417$\pm$0.035 & 0.377$\pm$0.008 & 0.384$\pm$0.016\\
 & 720 & 0.919 & 0.581 & 3.030 & 1.090 & 0.572 & 0.478 & 0.433$\pm$0.014 & 0.408$\pm$0.048 & 0.482$\pm$0.030 & 0.446$\pm$0.069 & \textbf{0.421$\pm$0.012} & \textbf{0.402$\pm$0.039} \\
 \midrule
 \multirow{ 4}{*}{\textbf{ETTm2
}} & 96 & 0.245 & 0.313 & 0.288 & 0.347 & 0.211 & 0.294 & \textbf{0.157$\pm$0.002} & 0.179$\pm$0.028 & 0.159$\pm$0.002 & 0.181$\pm$0.029 & 0.159$\pm$0.003 & \textbf{0.179$\pm$0.024}\\
 & 192 & 0.351 & 0.375 & 0.422 & 0.418 & 0.278 & 0.338 & 0.221$\pm$0.002 & 0.239$\pm$0.024 & 0.222$\pm$0.002 & 0.246$\pm$0.033 & \textbf{0.216$\pm$0.002} & \textbf{0.233$\pm$0.022} \\
 & 336 & 0.452 & 0.427 & 0.583 & 0.495 & 0.343 & 0.377 & 0.273$\pm$0.000 & 0.305$\pm$0.046 & 0.283$\pm$0.007 & 0.312$\pm$0.048 & \textbf{0.271$\pm$0.006} & \textbf{0.298$\pm$0.035} \\
 & 720 & 0.554 & 0.480 & 0.918 & 0.625 & 0.459 & 0.441 & 0.360$\pm$0.007 & 0.318$\pm$0.052 & 0.373$\pm$0.010 & 0.328$\pm$0.059 & \textbf{0.352$\pm$0.005} & \textbf{0.317$\pm$0.053} \\
 \midrule
  \multirow{ 4}{*}{\textbf{Weather
}} & 96 &0.197 & 0.256 & 0.315 & 0.295 & 0.188 & 0.248 & 0.141$\pm$0.001 & 0.158$\pm$0.026 & 0.147$\pm$0.002 & 0.165$\pm$0.024 & \textbf{0.140$\pm$0.002} & \textbf{0.157$\pm$0.023} \\
 & 192 & 0.292 & 0.324 & 0.416 & 0.353 & 0.265 & 0.309 & 0.189$\pm$0.003 & 0.207$\pm$0.023 & 0.191$\pm$0.002 & 0.211$\pm$0.031 & \textbf{0.187$\pm$0.003} & \textbf{0.204$\pm$0.026} \\
 & 336 & 0.433 & 0.390 & 0.576 & 0.426 & 0.409 & 0.374 & \textbf{0.240$\pm$0.002} & \textbf{0.252$\pm$0.018} & 0.246$\pm$0.006 & 0.260$\pm$0.025 & 0.242$\pm$0.004 & 0.255$\pm$0.021\\
 & 720 & 0.633 & 0.483 & 1.035 & 0.576 & 0.612 & 0.460 & \textbf{0.311$\pm$0.008} & \textbf{0.319$\pm$0.012} & 0.322$\pm$0.006 & 0.332$\pm$0.009 & 0.316$\pm$0.008 & 0.323$\pm$0.018\\
 \midrule
    \multirow{ 4}{*}{\textbf{Electricity
}} & 96 & 0.161 & 0.259 & 2.006 & 1.077 & 0.164 & 0.260 & 0.145$\pm$0.002 & 0.178$\pm$0.046 & 0.164$\pm$0.010 & 0.198$\pm$0.041 & \textbf{0.143$\pm$0.001} & \textbf{0.176$\pm$0.046} \\
 & 192 & 0.180 & 0.275 & 2.399 & 1.160 & 0.181 & 0.276 & \textbf{0.159$\pm$0.002} & \textbf{0.191$\pm$0.047} & 0.173$\pm$0.005 & 0.205$\pm$0.049 & \textbf{0.159$\pm$0.002} & 0.192$\pm$0.045 \\
 & 336 & 0.203 & 0.295 & 3.151 & 1.304 & 0.202 & 0.296 & \textbf{0.172$\pm$0.002} & \textbf{0.204$\pm$0.044} & 0.191$\pm$0.003 & 0.222$\pm$0.046 & 0.173$\pm$0.001 & 0.207$\pm$0.047 \\
 & 720 & 0.268 & 0.342 & 4.664 & 1.592 & 0.260 & 0.340 & \textbf{0.203$\pm$0.003} & \textbf{0.235$\pm$0.044} & 0.233$\pm$0.001 & 0.264$\pm$0.044 & 0.206$\pm$0.002 & 0.238$\pm$0.046\\
\midrule
\end{tabular}
}
\label{tab:ett_zs_ft}
\end{table}

\FloatBarrier
\section{Probability Quantification}\label{sec:prob_forecast_appendix}

For fine-tuning Delphyne, we use a learning rate of 5e-5, dropout of 0.2, batch size of 128, and a linear warmup for the learning rate of 50 steps. For all datasets, we use a context length of 1000 except Walmart, for which we use 50-100. We use early-stopping based on the validation loss. Similarly, we stick to the default hyperparameters without additional searching.

For evaluation, we use CRPS \citep{gneiting2007strictly}, MSIS \citep{makridakis2020m4}, symmetric mean absolute percentage error (sMAPE) \citep{hyndman2014errors}, mean absolute scaled error (MASE) \citep{hyndman2006another}, normalized deviation (ND), and normalized root mean squared error (NRMSE) \citep{yu2016temporal}.

The CRPS \citep{gneiting2007strictly} is a probabilistic forecasting evaluation metric, given a forecasted distribution with c.d.f. \(F\) and ground truth \(y\), it is defined as: 
\begin{align*}
    \textrm{CRPS} & = \int_0^1 2 \Lambda_{\alpha}(F^{-1}(\alpha), y) d\alpha \\
    \Lambda_{\alpha}(q, y) & = (\alpha - \1_\mathrm{y < q})(y - q),
\end{align*}
where \(\Lambda_{\alpha}\) is the \(\alpha\)-quantile loss, also known as the pinball loss at quantile level \(\alpha\).
To compute a normalized metric, the mean weighted sum quantile loss \citep{park2022quantile}, defined as the average of \(K\) quantiles:
\begin{align*}
    \textrm{CRPS} & \approx \frac{1}{K} \sum_{k=1}^K \textrm{wQL}[\alpha_k] \\
    \textrm{wQL}[\alpha] & = 2 \frac{\sum_{t} \Lambda_{\alpha}(\hat{q}_{t}(\alpha), y_t)}{\sum_{t} |y_{t}|},
\end{align*}
where \(\hat{q}_t(\alpha)\) is the forecasted \(\alpha\)-quantile at time step \(t\). We take \(K = 9, \alpha_1 = 0.1, \alpha_2 = 0.2, \ldots, \alpha_9 = 0.9\) in practice.

The MSIS \citep{makridakis2020m4} is a metric to evaluate uncertainty around point forecasts. Given an upper bound forecast \(U_t\) (0.975 quantile) and lower bound forecast \(L_t\) (0.025 quantile) the MSIS is defined as:
\[
\textrm{MSIS} = \frac{1}{h} \frac{\sum_{t=1}^h (U_t - L_t) + \frac{2}{a}(L_t - Y_t) \mathbb{I}_{\{Y_t < L_t\}} + \frac{2}{a}(Y_t - U_t) \mathbb{I}_{\{Y_t > U_t\}}}{\frac{1}{n-m}\sum_{t=m+1}^n |Y_t - Y_{t-m}|}
\]
where \(a = 0.05\) is the significance level for a 95\% forecast interval, over a forecast horizon of length \(h\), and \(m\) is the seasonal factor.

\begin{table}[htbp]
\centering
\caption{Full results for probabilistic forecasting experiments for different Delphyne models. The best results are highlighted in \textbf{bold}. Fine-tuning results are averaged across three experimental runs, with reported mean$\pm$std. Fine-tuning improves the model performances across the board.}
\vspace{-0.13 in}
\resizebox{\textwidth}{!}{%
\begin{tabular}{lccccccc}
\toprule
& & \methodl-ZS & \methodf-ZS & \methoda-ZS & \methodl-FT & \methodf-FT & \methoda-FT  \\
\midrule
\multirow{5}{*}{\textbf{Electricity}} 
& \textbf{CRPS} & 0.153 & 0.362 & 0.159 & \textbf{0.135$\pm$0.007} & 0.136$\pm$0.003 & 0.140$\pm$0.005\\
& \textbf{MSIS} &30.006 & 36.197 & 29.293 & 27.106$\pm$1.357 &  24.326$\pm$1.498 &\textbf{21.820$\pm$1.383}\\
& \textbf{sMAPE} & 0.229 & 0.535 & 0.233   & 0.210$\pm$0.008 & 0.224$\pm$0.007 & 0.215$\pm$0.008\\
& \textbf{MASE} & 1.972 & 4.325 & 2.031 & \textbf{1.783$\pm$0.089} & 1.824$\pm$0.039 & 1.839$\pm$0.080\\
& \textbf{ND} & 0.177 & 0.453 & 0.182 & \textbf{0.158$\pm$0.008} & 0.163$\pm$0.004 & 0.164$\pm$0.005 \\
& \textbf{NRMSE } &1.037 & 2.915 & 1.084 & 0.948$\pm$0.034 & \textbf{0.936$\pm$0.027} & 0.986$\pm$0.007 \\
\midrule
\multirow{5}{*}{\textbf{Solar}} 
& \textbf{CRPS} & \textbf{0.893} & 1.126 & 0.905 & 1.223$\pm$0.104 & 1.423$\pm$0.092 & 1.306$\pm$0.103 \\
& \textbf{MSIS} & 2.518 & 3.279 & 2.733 & \textbf{1.600$\pm$0.409} & 3.085$\pm$0.859 & 2.029$\pm$0.520\\
& \textbf{sMAPE} & 1.650 & 1.696 & 1.650 & \textbf{1.482$\pm$0.005} & 1.706$\pm$0.003 & 1.499$\pm$0.009  \\
& \textbf{MASE} &  1.631 & 1.543 & 1.710 & \textbf{1.010$\pm$0.102} & 2.348$\pm$0.462 & 1.076$\pm$0.164\\
& \textbf{ND} & 0.249 & \textbf{0.233} & 0.263 & 0.245$\pm$0.016 & 0.356$\pm$0.070 & 0.290$\pm$0.028 \\
& \textbf{NRMSE } & 2.486 & 3.070 & 2.483 & 1.090$\pm$0.231 & 4.544$\pm$0.108 & \textbf{1.060$\pm$0.129} \\
\midrule
\multirow{5}{*}{\textbf{Walmart}} 
& \textbf{CRPS} & 0.096 & 0.106 & 0.093 & 0.091$\pm$0.001 & 0.103$\pm$0.001 & \textbf{0.083$\pm$0.001} \\
& \textbf{MSIS} & 4.709 & 5.284 & 4.741 & \textbf{4.203$\pm$0.072} & 4.518$\pm$0.165 & 4.559$\pm$0.289\\
& \textbf{sMAPE} & 0.185 & 0.201 & 0.184  & 0.079$\pm$0.003 & 0.193$\pm$0.006 & \textbf{0.088$\pm$0.003} \\
& \textbf{MASE} & 0.666 & 0.703 & 0.645 & 0.669$\pm$0.017 & 0.719$\pm$0.023 & \textbf{0.660$\pm$0.002} \\
& \textbf{ND} & 0.132 & 0.142 & 0.126 & 0.107$\pm$0.002 & 0.141$\pm$0.004 & \textbf{0.101$\pm$0.001} \\
& \textbf{NRMSE } & 0.291 & 0.303 & 0.270 & \textbf{0.260$\pm$0.001 }& 0.325$\pm$0.016 & 0.279$\pm$0.003 \\
\midrule
\multirow{5}{*}{\textbf{Weather}} 
& \textbf{CRPS} & 0.061 & 0.074 & 0.064 & \textbf{0.041$\pm$0.003} & 0.045$\pm$0.004 & 0.042$\pm$0.005 \\
& \textbf{MSIS} & 6.012 & 8.331 & 6.080 & 4.600$\pm$0.100 & \textbf{4.304$\pm$0.092} & 4.467$\pm$0.053\\
& \textbf{sMAPE} & 0.959 & 0.812 & 0.906  & 0.895$\pm$0.236 & 1.059$\pm$0.178 & \textbf{0.890$\pm$0.214} \\
& \textbf{MASE} & 0.711 & 0.834 & 0.701 & 0.510$\pm$0.050 & 0.535$\pm$0.020 & \textbf{0.505$\pm$0.058} \\
& \textbf{ND} & 0.088 & 0.108 & 0.095 & \textbf{0.056$\pm$0.008} & 0.064$\pm$0.012 & 0.057$\pm$0.012 \\
& \textbf{NRMSE } & 0.255 & 0.403 & 0.270 & \textbf{0.209$\pm$0.007} & 0.218$\pm$0.016 & 0.212$\pm$0.015
\\
\midrule
\multirow{5}{*}{\textbf{Istanbul Traffic}} 
& \textbf{CRPS} & 0.155 & 0.468 & \textbf{0.149} & 0.209$\pm$0.007 & 0.218$\pm$0.016 & 0.212$\pm$0.015
\\
& \textbf{MSIS} & 11.656 & 15.971 & 9.989 & 5.634$\pm$2.351  & 5.454$\pm$1.433 & \textbf{4.328$\pm$0.536} \\
& \textbf{sMAPE} & 0.724 & 5.171 & 0.352 & \textbf{0.236$\pm$0.018} & 0.249$\pm$0.021 & 0.242$\pm$0.017 \\
& \textbf{MASE} & 0.799 & 2.790 & 0.772 & 0.580$\pm$0.074 & 0.611$\pm$0.063 & \textbf{0.558$\pm$0.018} \\
& \textbf{ND} & 0.181 & 0.632 & 0.175 & 0.131$\pm$0.017 & 0.139$\pm$0.014&  \textbf{0.127$\pm$0.004}  \\
& \textbf{NRMSE } & 0.284 & 0.755 & 0.273  & 0.220$\pm$0.031 & 0.219$\pm$0.025&  \textbf{0.189$\pm$0.010} \\
\midrule
\multirow{5}{*}{\textbf{Turkey Power}} 
& \textbf{CRPS} & 0.046 & 0.148 & 0.046 & 0.035$\pm$0.002 & 0.062$\pm$0.002 & \textbf{0.035$\pm$0.001} \\
& \textbf{MSIS} & 6.299 & 19.902 & 6.269 & 5.462$\pm$0.117 & 7.856$\pm$0.197 & \textbf{5.384$\pm$0.346} \\
& \textbf{sMAPE} & 0.172 & 0.277 & 0.176 & \textbf{0.167$\pm$0.001} & 0.195$\pm$0.002 & 0.168$\pm$0.002\\
& \textbf{MASE} & 0.886 & 1.753 & 0.891 & 0.794$\pm$0.004 & 1.009$\pm$0.015 & \textbf{0.790$\pm$0.018} \\
& \textbf{ND} & 0.058 & 0.187 & 0.059 & 0.045$\pm$0.002 & 0.080$\pm$0.002 & \textbf{0.045$\pm$0.001} \\
& \textbf{NRMSE } & 0.132 & 0.486 & 0.132 & \textbf{0.096$\pm$0.006} & 0.202$\pm$0.010 & 0.098$\pm$0.003 \\
\bottomrule
\end{tabular}%
}
\end{table}

\begin{table}[htbp]
\centering
\caption{Full results for probabilistic forecasting experiments. The best results are highlighted in \textbf{bold}, and the second best results are \underline{underlined}. (The baseline results are taken from \cite{woo2024unifiedtraininguniversaltime}.)}
\label{tab:prob_forecast_full}
\vspace{-0.12 in}
\resizebox{1.0\textwidth}{!}{%
\begin{tabular}{lccccccccccc}
\toprule
& & \multicolumn{2}{c}{\textbf{Zero-shot}} & \multicolumn{1}{c}{\textbf{Finetuned}} & \multicolumn{4}{c}{\textbf{Full-shot}} & \multicolumn{2}{c}{\textbf{Baseline}}\\ 
\cmidrule(lr){3-4} \cmidrule(lr){5-5} \cmidrule(lr){6-9} \cmidrule(lr){10-11}
& & \methoda-ZS & MOIRAI & \methoda-FT &  PatchTST  & TiDE & TFT & DeepAR & AutoARIMA & Seasonal Naive   \\
\midrule
\multirow{5}{*}{\textbf{Electricity}} 
& \textbf{CRPS} &  0.159 & 0.055 & 0.140$\pm$0.005 &  0.052$\pm$0.00 & \textbf{0.048$\pm$0.00} & \underline{0.050$\pm$0.00} & 0.065$\pm$0.01 & 0.327 & 0.070 \\
& \textbf{MSIS} & 29.293 & 6.172 & 21.820$\pm$1.383 & \underline{5.744$\pm$0.12} & \textbf{5.672$\pm$0.08} & 6.278$\pm$0.24 & 6.893$\pm$0.82 & 29.412 & 35.251 \\ 
& \textbf{sMAPE} & 0.233 & 0.111 & 0.215$\pm$0.008 & 0.107$\pm$0.00 & \textbf{0.102$\pm$0.00} & \underline{0.106$\pm$0.01} & 0.118$\pm$0.02 & 0.318 & 0.108 \\
& \textbf{MASE} &  2.031 & 0.792 & 1.839$\pm$0.080 & 0.753$\pm$0.01 & \underline{0.706$\pm$0.02} & \underline{0.747$\pm$0.03} & 0.844$\pm$0.16 & 3.229 & 0.881 \\
& \textbf{ND} & 0.182 & 0.069 & 0.164$\pm$0.005 &  0.065$\pm$0.00 & \textbf{0.061$\pm$0.00} & \underline{0.063$\pm$0.00} & 0.080$\pm$0.02 & 0.357 & 0.070 \\
& \textbf{NRMSE } & 1.084 & 0.551 & 0.986$\pm$0.007 & \underline{0.506$\pm$0.02} & 0.514$\pm$0.02 & 0.511$\pm$0.02 & 0.704$\pm$0.11 & 3.296 & \textbf{0.478} \\
\midrule
\multirow{5}{*}{\textbf{Solar}} 
& \textbf{CRPS} & 0.905 & 0.419 & 1.306$\pm$0.103 & 0.518$\pm$0.09 & \textbf{0.420$\pm$0.00} & 0.446$\pm$0.03 & \underline{0.431$\pm$0.01} & 1.055 & 0.512 \\
& \textbf{MSIS} & \underline{2.733} & 7.011 & \textbf{2.029$\pm$0.520}  & 8.447$\pm$1.59 & 13.754$\pm$0.32 & 8.057$\pm$3.51 & 11.181$\pm$0.67 & 25.849 & 48.130 \\
& \textbf{sMAPE} & 1.650 & 1.410 & 1.499$\pm$0.009 &  1.501$\pm$0.10 & 1.400$\pm$0.00 & 1.391$\pm$0.01 & \underline{1.385$\pm$0.00} & 1.685 & \textbf{0.691} \\
& \textbf{MASE} & 1.710 & 1.292 & \textbf{1.076$\pm$0.164} & 1.607$\pm$0.25 & 1.265$\pm$0.02 & 1.399$\pm$0.11 & 1.222$\pm$0.01 & 2.583 & \underline{1.203} \\
& \textbf{ND} &  \textbf{0.263} & 0.551 & \underline{0.290$\pm$0.028} & 0.685$\pm$0.11 & 0.538$\pm$0.01 & 0.594$\pm$0.05 & 0.520$\pm$0.00 & 1.098 & 0.512 \\ 
& \textbf{NRMSE } &  2.483 & \underline{1.034} & 1.060$\pm$0.129 & 1.408$\pm$0.26 & 1.093$\pm$0.00 & 1.236$\pm$0.06 & \textbf{1.033$\pm$0.01} & 1.784 & 1.168 \\
\midrule
\multirow{5}{*}{\textbf{Walmart}} 
& \textbf{CRPS} &  0.093 & 0.093 & 0.083$\pm$0.001 &  \underline{0.082$\pm$0.01} & \textbf{0.077$\pm$0.00} & 0.087$\pm$0.00 & 0.121$\pm$0.00 & 0.124 & 0.151 \\
& \textbf{MSIS} & \underline{4.741} & 8.421 & \textbf{4.559$\pm$0.289} & 6.005$\pm$0.21 & 6.258$\pm$0.12 & 8.718$\pm$0.10 & 12.502$\pm$0.03 & 9.888 & 49.458 \\
& \textbf{sMAPE} & 0.184 & 0.168 & \textbf{0.088$\pm$0.003} & 0.150$\pm$0.01 & \underline{0.145$\pm$0.00} & 0.172$\pm$0.00 & 0.216$\pm$0.00 & 0.219 & 0.205 \\
& \textbf{MASE} &  \textbf{0.645} & 0.964 & \underline{0.660$\pm$0.002} & 0.867$\pm$0.09 & 0.814$\pm$0.01 & 0.948$\pm$0.02 & 1.193$\pm$0.02 & 1.131 & 1.236 \\
& \textbf{ND} & 0.126 & 0.117 & \underline{0.101$\pm$0.001} & 0.105$\pm$0.01 & \textbf{0.097$\pm$0.00} & 0.108$\pm$0.00 & 0.147$\pm$0.00 & 0.141 & 0.151 \\
& \textbf{NRMSE } & 0.270 & 0.291 & 0.279$\pm$0.003 & \underline{0.218$\pm$0.02} & \textbf{0.204$\pm$0.00} & 0.235$\pm$0.01 & 0.298$\pm$0.00 & 0.305 & 0.328 \\
\midrule
\multirow{5}{*}{\textbf{Weather}} 
& \textbf{CRPS} & 0.064 & \textbf{0.041} & \underline{0.042$\pm$0.005} & 0.059$\pm$0.01 & 0.054$\pm$0.00 & 0.043$\pm$0.00 & 0.132$\pm$0.01 & 0.252 & 0.068 \\
& \textbf{MSIS} & 6.080 &  \underline{5.136} & \textbf{4.467$\pm$0.053} & 7.759$\pm$0.49 & 8.095$\pm$1.74 & 7.791$\pm$0.44 & 21.651$\pm$17.34 & 19.805 & 31.293 \\
& \textbf{sMAPE} & 0.906  & \underline{0.623} &  0.890$\pm$0.214  & 0.668$\pm$0.01 & 0.636$\pm$0.01 & 0.672$\pm$0.01 & 0.776$\pm$0.05 & 0.770 & \textbf{0.401} \\
& \textbf{MASE} & 0.701 & \textbf{0.487} & \underline{0.505$\pm$0.058} & 0.844$\pm$0.19 & 0.832$\pm$0.13 & 0.692$\pm$0.02 & 3.170$\pm$3.47 & 0.938 & 0.782 \\ 
& \textbf{ND} &  0.095 & \textbf{0.048} & 0.057$\pm$0.012 & 0.072$\pm$0.02 & 0.066$\pm$0.01 & \underline{0.051$\pm$0.00} & 0.163$\pm$0.15 & 0.139 & 0.068\\
& \textbf{NRMSE } & 0.270 & 0.417 & \underline{0.212$\pm$0.015} & 0.260$\pm$0.01 & 0.214$\pm$0.00 & \underline{0.211$\pm$0.00} & 0.486$\pm$0.43 & 0.465 & 0.290 \\
\midrule
\multirow{5}{*}{\textbf{Istanbul Traffic}} 
& \textbf{CRPS} & 0.149 & 0.116& 0.212$\pm$0.015 & 0.112$\pm$0.00 & 0.110$\pm$0.01 & \underline{0.110$\pm$0.01} & \textbf{0.108$\pm$0.00} & 0.589 & 0.257 \\
& \textbf{MSIS} & 9.989 & 4.461 & 4.328$\pm$0.536 & \textbf{3.813$\pm$0.09} & 4.752$\pm$0.17 & \underline{4.057$\pm$0.44} & 4.094$\pm$0.31 & 16.317 & 45.473 \\
& \textbf{sMAPE} &  0.352 & 0.284 & \textbf{0.242$\pm$0.017} & 0.287$\pm$0.01 & 0.280$\pm$0.01 & 0.287$\pm$0.01 & \underline{0.249$\pm$0.01} & 1.141 & 0.391 \\
& \textbf{MASE} &  0.772 & 0.644 & \textbf{0.558$\pm$0.018} & 0.653$\pm$0.02 & 0.618$\pm$0.03 & 0.620$\pm$0.03 & \underline{0.613$\pm$0.03} & 3.358 & 1.137 \\
& \textbf{ND} &  0.175 & 0.146 & \textbf{0.127$\pm$0.004} & 0.148$\pm$0.01 & 0.140$\pm$0.01 & 0.141$\pm$0.01 & \underline{0.139$\pm$0.01} & 0.758 & 0.257 \\
& \textbf{NRMSE } & 0.273 & 0.194 & 0.189$\pm$0.010 & 0.190$\pm$0.01 & \underline{0.185$\pm$0.01} & 0.185$\pm$0.01 & \textbf{0.181$\pm$0.01} & 0.959 & 0.384 \\
\midrule
\multirow{5}{*}{\textbf{Turkey Power}} 
& \textbf{CRPS} & 0.046 & \underline{0.040} & \textbf{0.035$\pm$0.001} & 0.054$\pm$0.01 & 0.046$\pm$0.01 & 0.039$\pm$0.00 & 0.066$\pm$0.02 & 0.116 & 0.085 \\
& \textbf{MSIS} & \underline{6.269} & 6.766 & \textbf{5.384$\pm$0.346} & 8.978$\pm$0.51 & 8.579$\pm$0.52 & 7.943$\pm$0.31 & 13.520$\pm$1.17 & 14.863 & 36.256 \\
& \textbf{sMAPE} &  0.176 & 0.378 & \underline{0.168$\pm$0.002} &  0.416$\pm$0.01 & 0.389$\pm$0.00 & 0.383$\pm$0.00 & 0.404$\pm$0.01 & 0.244 & \textbf{0.125} \\
& \textbf{MASE} &  0.891 & \underline{0.888} & \textbf{0.790$\pm$0.018} & 1.234$\pm$0.12 & 0.904$\pm$0.02 & 0.890$\pm$0.05 & 1.395$\pm$0.30 & 1.700 & 0.906 \\
& \textbf{ND} & 0.059 & 0.051 &  \textbf{0.045$\pm$0.001} & 0.071$\pm$0.01 & 0.059$\pm$0.01 & \underline{0.049$\pm$0.00} & 0.083$\pm$0.02 & 0.150 & 0.085 \\
& \textbf{NRMSE } & 0.132 & 0.118& \textbf{0.098$\pm$0.003} & 0.158$\pm$0.01 & 0.139$\pm$0.03 & \underline{0.104$\pm$0.01} & 0.181$\pm$0.05 & 0.383 & 0.231 \\
\bottomrule
\end{tabular}%
}
\end{table}

\section{Anomaly Detection}
\label{sec:anomaly_detection_description}

\subsection{Anomaly Detection Experiment Setup}

Our experimental setup is similar to that of \citet{goswami2024moment}. Following \citet{goswami2023unsupervised}, we used a fixed anomaly detection window size of 512 and downsampled all time-series datasets longer than 2560 timesteps by a factor of 10 to speed up the training and
evaluation process. We use the mean squared error between forecasts and observations as the anomaly criterion. We get forecasts from our model by masking out nonoverlapping patches of 32 from the window of 512. Even though Delphyne was pre-trained to forecast, we noticed that Delphyne was able impute values in other parts of each time series. For training, to improve imputation performance, each row of the dataset is to forecast a random patches of size 32, not just at the end of the time series.

\subsection{Full Comparison Results}
Table \ref{tab:anomaly_detection_results_appendix} shows the full adjusted F1 results across UCR Anomaly Archive.
\begin{table}[!tbh]
\centering
\resizebox{\linewidth}{!}{
\begin{tabular}{r|cccccccccccc}
\toprule
\textbf{Model name} & \textbf{Anomaly Transformer}  & \textbf{MOMENT} & \textbf{DGHL} & \textbf{GPT4TS} & \textbf{TimesNet} & AnomalyTransformer & \methoda-ZS & \methoda-FT\\
\midrule
InternalBleeding4 & NaN &  NaN & NaN & NaN & NaN & NaN & 0.717 & 0.996 \\
1sddb40 & 0.030 & 0.540 & 0.390 & 0.190 & 0.680 & 0.640 & 0.818 & 0.754 \\
BIDMC1 & 0.990  & 1.000 & 1.000 & 1.000 & 1.000 & 0.690 & 0.390 & 0.899\\
CHARISfive & 0.010 & 0.130 & 0.020 & 0.020 & 0.080 & 0.360 & 0.017 & 0.015 \\
CHARISten & 0.020 & 0.110 & 0.040 & 0.100 & 0.030 & 0.430 & 0.034 & 0.040 \\
CIMIS44AirTemperature3 & 0.060  & 0.980 & 0.500 & 0.180 & 0.470 & 0.640 & 0.167 & 1.000 \\
CIMIS44AirTemperature5 & 0.390  & 0.990 & 0.960 & 0.200 & 0.710 & 0.780 & 0.225 & 1.000 \\
ECG2 & 1.000  & 1.000 & 0.620 & 0.900 & 1.000 & 0.830 & 0.864 & 0.772 \\
ECG3 & 0.360  & 0.980 & 0.800 & 0.840 & 0.480 & 0.540 & 0.142 & 0.727\\
Fantasia & 0.750  & 0.950 & 0.660 & 0.870 & 0.550 & 0.730 & 0.882 & 0.833 \\
GP711MarkerLFM5z4 & 0.930  & 1.000 & 0.500 & 0.640 & 0.950 & 0.540 & 0.837 & 1.000\\
GP711MarkerLFM5z5 & 0.760  & 0.970 & 0.310 & 0.480 & 0.900 & 0.690 & 0.717 & 1.000 \\
InternalBleeding5 & 0.940  & 1.000 & 1.000 & 0.920 & 1.000 & 0.460 & 0.883 & 0.914 \\
Italianpowerdemand & 0.010 & 0.740 & 0.590 & 0.010 & 0.440 & 0.450 & 0.087 & 0.259 \\
Lab2Cmac011215EPG5 & 0.990 & 0.980 & 0.340 & 0.600 & 0.990 & 0.770 & 0.477 & 0.672 \\
Lab2Cmac011215EPG6 & 0.410  & 0.100 & 0.260 & 0.100 & 0.170 & 0.700 & 0.118 & 0.209 \\
MesoplodonDensirostris & 1.000 & 0.840 & 0.790 & 1.000 & 1.000 & 0.850 & 0.532 & 0.947 \\
PowerDemand1 & 0.870 & 0.440 & 0.490 & 0.760 & 0.950 & 0.720 & 0.433 & 0.810 \\
TkeepFirstMARS & 0.010  & 0.150 & 0.020 & 0.020 & 0.230 & 0.520 & 0.018 & 0.061\\
TkeepSecondMARS & 0.830 & 1.000 & 0.160 & 0.120 & 0.950 & 0.720 & 0.057 & 0.625 \\
WalkingAceleration5 & 0.990  & 1.000 & 0.910 & 0.870 & 0.930 & 0.940 & 0.634 & 0.843 \\
apneaecg & 0.400 & 0.200 & 0.250 & 0.310 & 0.260 & 0.580 & 1.000 & 1.000 \\
apneaecg2 & 0.650 & 1.000 & 1.000 & 1.000 & 0.650 & 0.790 & 0.213 & 0.213\\
gait1 & 0.180  & 0.360 & 0.070 & 0.410 & 0.520 & 0.630 & 0.204 & 0.144 \\
gaitHunt1 & 0.080 & 0.430 & 0.020 & 0.100 & 0.300 & 0.810 & 0.008 & 0.007 \\
insectEPG2 & 0.120  & 0.230 & 0.140 & 0.810 & 0.960 & 0.650 & 0.093 & 0.385 \\
insectEPG4 & 0.980  & 1.000 & 0.460 & 0.210 & 0.850 & 0.690 & 0.068 & 0.840 \\
ltstdbs30791AS & 1.000 & 1.000 & 1.000 & 1.000 & 1.000 & 0.780 & 0.080 & 0.120 \\
mit14046longtermecg & 0.450  & 0.590 & 0.530 & 0.580 & 0.600 & 0.790 & 0.939 & 0.939\\
park3m & 0.150  & 0.640 & 0.200 & 0.630 & 0.930 & 0.630 & 0.232 & 0.753 \\
qtdbSel1005V & 0.410 & 0.650 & 0.400 & 0.390 & 0.530 & 0.520 & 0.494 & 0.412 \\
qtdbSel100MLII & 0.420  & 0.840 & 0.410 & 0.600 & 0.870 & 0.620 & 0.417 & 0.402 \\
resperation1 & 0.000& 0.150 & 0.030 & 0.010 & 0.030 & 0.750 & 0.006 & 0.006 \\
s20101mML2 & 0.690 & 0.710 & 0.150 & 0.050 & 0.080 & 0.640 & 1.000 & 1.000 \\
sddb49 & 0.890 & 1.000 & 0.880 & 0.940 & 1.000 & 0.660 & 0.781 & 0.820 \\
sel840mECG1 & 0.160  & 0.660 & 0.280 & 0.210 & 0.360 & 0.620 & 0.247 & 0.235 \\
sel840mECG2 & 0.150  & 0.390 & 0.320 & 0.280 & 0.210 & 0.590 & 0.484 & 0.507 \\
tilt12744mtable & 0.070 & 0.240 & 0.100 & 0.000 & 0.030 & 0.480 & 0.031 & 0.080 \\
tilt12754table & 0.230 & 0.640 & 0.040 & 0.060 & 0.050 & 0.600 & 0.017 & 0.084 \\
tiltAPB2 & 0.920  & 0.980 & 0.360 & 0.830 & 0.380 & 0.770 & 0.416 & 0.692 \\
tiltAPB3 & 0.170  & 0.850 & 0.030 & 0.050 & 0.090 & 0.680 & 0.029 & 0.068\\
weallwalk & 0.000 & 0.580 & 0.070 & 0.130 & 0.170 & 0.730 & 0.041 & 0.667 \\
\bottomrule
\end{tabular}}
\caption{Anomaly detection performance measured using adj. best $F_1$ for a subset of 45 datasets sampled from the UCR Anomaly archive.}
\label{tab:anomaly_detection_results_appendix}
\end{table}

\FloatBarrier
\section{Negative Transfer}
\label{sec:negative_transfer}

For all our synthetic data experiments, we train with 8 layers, the attention is of 1024 dimension shared between 8 heads,
following to a maximum width of 4098. We used no dropout. The model is
trained on negative log likelihood loss of a Gaussian. The model was trained on 100K steps with a fixed patch
size of 1. For optimization, we used a batch size of 64 and
employed the AdamW optimizer with the following hyperparameters: lr $= 1e-4$, weight
decay $= 1e-5$, $\beta_1$ = 0.9, and $\beta_2$ = 0.98. A learning rate scheduler was applied, featuring linear
warmup for the first 5,000 steps, followed by cosine annealing down to 1e-5. 

For our synthetic data, we use wavelet functions:
\begin{align*}
x_t &= (d * t / T - c)\ \text{sin}\left(a * t + b \right) + \epsilon_t
\\ \epsilon_t &\sim \mathcal{N}(0, 0.2)
\end{align*}
where the parameters are $\{a, b, c, d\}$ and $T$ is the number of time steps of the time series, and GARCH:
\begin{align*}
x_t &= \mu + \epsilon_t
\\ \epsilon_t &\sim \mathcal{N}(0, \sigma_t)
\\ \sigma^2_t &= \omega + a x_{t-1}^2 + b \epsilon_{t-1}^2
\end{align*}
where the parameters are $\{\omega, a, b, \sigma_0, \mu\}$.

\subsection{Pre-training with GARCH and Wavelet Data}
\label{ssec:pretrain_with_garch}

For generating GARCH data, we use:
\begin{align*}
    \mu &= 0
\\ \sigma_0 &= \omega \sim \text{U}(0,1 )
\\ a & \sim \text{U}(0, 1)
\\ b & \sim \text{U}(0, 1 - a)
\end{align*}

For generating wavelet data, we use:
\begin{align*}
    a & \sim \{0.1, 0.2, 0.6,0.8 \}
\\ b & \sim \{0,5,10\}
\\ c & \sim \{0.0, 0.3, 0.6, 0.9 \}
\\ d & \sim \{0.5, 0.9 \}
\end{align*}
where the sets denote a uniform sample from those choices.

\subsection{Bayesian MCMC}
\label{ssec:bayesian_mcmc}

For the wavelet distribution, we use:
\begin{align*}
    T &= 32
\\ a & \sim \text{U}(0, 1)
\\ b & \sim \text{U}(0, 1)
\\ c & \sim \text{U}(0, 1)
\\ d & \sim \text{U}(0, 1)
\end{align*}
where $\text{U}$  denotes a Uniform distribution.

For the GARCH distribution, we use:
\begin{align*}
    \mu &= 0
\\ \sigma_0 &= \omega = 0.3^2
\\ a & \sim \text{U}(0, 0.2)
\\ b & \sim \text{U}(0, 0.2)
\end{align*}

For computing the NLL and the mean, we need to find the probability that a time-series comes from each distribution. For this, we computed the log-likelihood through 10K samples from the prior. For computing the posterior of the parameters given the data for each model, we use the NUTS sampler \citep{Hoffman2015nutssampler}.

\section{Additional Ablation Studies}

For pre-training, we use the same setup as in Section \ref{sec:negative_transfer}. For fine-tuning, similar to fine-tuning Delphyne, we early-stop based on a validation set.

\subsection{Context Length}
\label{ssec:ablation_context_len}

For these experiments, we use the same wavelet data generation as in Section \ref{ssec:pretrain_with_garch}. We used a finite set of configurations to test how well the model is able to create features specific to a dataset. The intuition is that with a smaller context length, the pre-training does not create features specific to the type of wavelet that generated the data but longer context lengths do. 

\subsubsection{Architecture for Context Lengths}
\textbf{Small} The small model is a transformer encoder of 6 encoder layers, a context length of 128, and a hidden dimension size of 512. It uses 8 attention heads, a feedforward dimension of 2048, and applies a GELU activation function. The model is designed to output attention weights, and features a dropout rate of 0.1 to prevent overfitting. It is trained using the Adam optimizer with a learning rate of 1e-4, betas of 0.9 and 0.98, and a weight decay of 1e-5. The configuration includes training, validation, and test batch sizes of 128, with warmup steps for the learning rate set at 5,000 out of 100,000 total training steps.

\textbf{Medium} The medium model contains 8 encoder layers and a context length of 128, featuring a hidden dimension size of 1024. The model uses 8 attention heads and a feedforward dimension of 4098 with GELU activation, while a dropout rate of 0.1 is applied for regularization. The model outputs attention weights and employs batch sizes of 64 for training, validation, and testing. It uses the Adam optimizer with a learning rate of 1e-4, betas of 0.9 and 0.98, a weight decay of 1e-5, and includes 5,000 warmup steps in a total of 100,000 training steps. 
\subsection{Masking Ratio}
\label{ssec:ablation_masking_ratio}

For these experiments, we use the same wavelet data generation as in Section \ref{ssec:pretrain_with_garch}.

\subsection{Multivariate}
\label{ssec: multivariate}

For these experiments, we use the same wavelet data generation as in Section \ref{ssec:pretrain_with_garch}. The main difference is each sample is four time series. We model two scenarios: one where the Wavelet data across rows are correlated, and another where they are uncorrelated. In the correlated scenario, the time-series data is generated using the same Wavelet function, differing only by additive Gaussian noise. In the uncorrelated scenario, the data is generated using different Wavelet functions.

\subsection{Output Distribution}
\label{ssec:output_distribution_ablation}

We use the same hyperparameter configuration for training \methoda, on three different output distributions: (1) Single Student T, (2) a mixture of Student-T distributions, and (3) a mixture of Normal, Student's-T, Log-normal, and negative binomial distributions. For every 10,000 training steps, we finetune the models on stock NLL task in the experiment section. The overall results are shown in Fig. \ref{fig:distributions}.

\section{List of Popular Models}
\label{sec:list_of_popular_models}
\begin{table*}[ht]
\centering
\caption{Comparison of Pre-trained Time-series Model}
\scalebox{0.6}{
\begin{tabular}{m{3cm}m{2cm}m{2cm}m{2cm}m{2cm}m{2cm}m{2cm}m{2cm}m{2cm}}
\toprule
\textbf{Feature} & \textbf{MOMENT} & \textbf{MOIRAI} & \textbf{Lag-Llama} & \textbf{Chronos} & \textbf{TimesFM} & \textbf{TimeGPT-1} & \textbf{TTM} & \textbf{Delphyne}   \\ 
& \citep{goswami2024moment} & \citep{woo2024unifiedtraininguniversaltime} & \citep{rasul2024lagllamafoundationmodelsprobabilistic} & \citep{ansari2024chronos} & \citep{das2024timesfm} & \citep{Garza2023TimeGPT1} & \citep{ekambaram2024tinytimemixersttms} & (This paper) \\\hline
Base Architecture & T5 encoder & Encoder-only transformer & Llama & T5 (encoder-decoder) & Decoder-only & Transformer & MLP-Mixer& Encoder-only transformer\\ \hline
Evaluation Tasks & Forecasting, Classification, Anomaly detection, Imputation & Forecasting & Forecasting & Forecasting & Forecasting & Forecasting & Forecasting & Forecasting, Anomaly detection \\ \hline
Tokenization & Fixed-length patches & Multi-scale Patches & Lag features & Scaling, Quantization & Fixed-length patches & ?  & Adaptive Patching & Fixed-length patches \\ \hline
Objective & Reconstruction Error & Forecast NLL of mixed-distributions & NLL of Student's t distribution & Cross-entropy loss & Forecast Error & ? & Forecasting Error & Forecast NLL of mixture of Student T's distributions \\ \hline
Distribution Prediction / Uncertainty Quantification &  & \checkmark & \checkmark & \checkmark &  & \checkmark(post hoc) & & \checkmark\\ \hline
Multivariates? & \checkmark (Anyvariate attention w. Channel independence) & \checkmark (Anyvariate attention + Flattening) &  & \checkmark &  & \checkmark (?) & (Channel Independence + Mixing) & \checkmark (Anyvariate attention +Flattening) \\ \hline
Context length & 512 & 1000--5000 & 1024 & 512 & 512 & ? &  512 & 512 x 32 \\ 
\bottomrule
\end{tabular}
\label{tab:comparison_of_pretrained_models}
}
\end{table*}

We provide a full table of the foundation models in Table \ref{tab:comparison_of_pretrained_models}. We compare popular models between 2022-2024: MOMENT \citep{goswami2024moment}, MOIRAI \citep{wang2024subgraphpoolingtacklingnegative}, Lag-Llama \citep{rasul2024lagllamafoundationmodelsprobabilistic}, Chronos \citep{ansari2024chronos}, TimesFM \citep{das2024timesfm}, TimeGPT-1 \citep{Garza2023TimeGPT1}, TTM \citep{ekambaram2024tinytimemixersttms}.

Among these popular models, only MOIRAI, Lag-Llama and TimeGPT-1 are able to provide output distributions and uncertainty quantifications. Specifically, Lag-Llama utilizes a single Student's T distribution which is less ideal to model asymmetries in forecasts, which is shown in our previous experiment. TimeGPT-1 uses a categorical output distribution. While it may potentially model any multi-modal distributions, the output distribution is tied to TimeGPT-1's language model architecture and training objective, offering less flexibility overall. Delphyne utilizes a mixture of Student's T distributions, which are simpler and more stable, as shown in our previous study. 

Many existing time-series foundation models excel in modeling single variates, which ignore the potential dependencies between variates (for example, when modeling US stock returns, many stocks in the same sectors are inter-correlated). We use the same any-variate attention mechanism as MOIRAI; we demonstrate in the previous section that any-variate attention performs reasonably well when both the variates are strongly or weakly correlated.

While many pre-trained time-series model aim to adapt to different forecast lengths, TTM has fixed forecast lengths. Its public model has a maximum context length of 1024 and a forecast length of 96, which is limited for various financial tasks. We argue that a good pre-trained time-series model should be agnostic to downstream tasks' forecast lengths and number of variates. In this context, Delphyne offers more flexibility.

Many popular time-series models, such as MOIRAI, Lag-Llama, and TTM, employ various patch sizes or use additional frequency information to capture different frequencies within datasets. We argue that these different patching methods aim to address the negative transfer effect across datasets. Since datasets across domains are collected at varying frequencies, these models leverage frequency information to create distinct embeddings for data at different granularities. 
In contrast, we believe that fine-tuning, despite being a post-hoc solution, offers the most effective means of mitigating the negative transfer effect.

\newpage
\FloatBarrier
\section{Visualizations}

\begin{figure}[!tb]
\begin{center}
\includegraphics[width=0.5\textwidth]{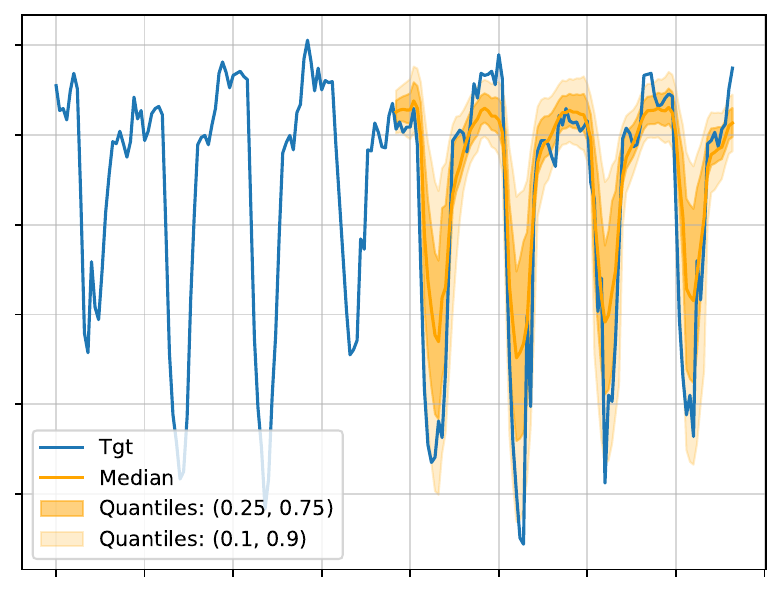}
\caption{Visualization of fine-tuned forecasts from \methoda on ETTh1 dataset. The quantiles represented are 0.1, 0.25, 0.5, 0.75, and 0.9.}
\label{fig:visualization_etth1}
\end{center}

\end{figure}

\begin{figure}[ht]
\centering
\begin{minipage}{0.45\textwidth}
    \centering
    \includegraphics[width=\textwidth]{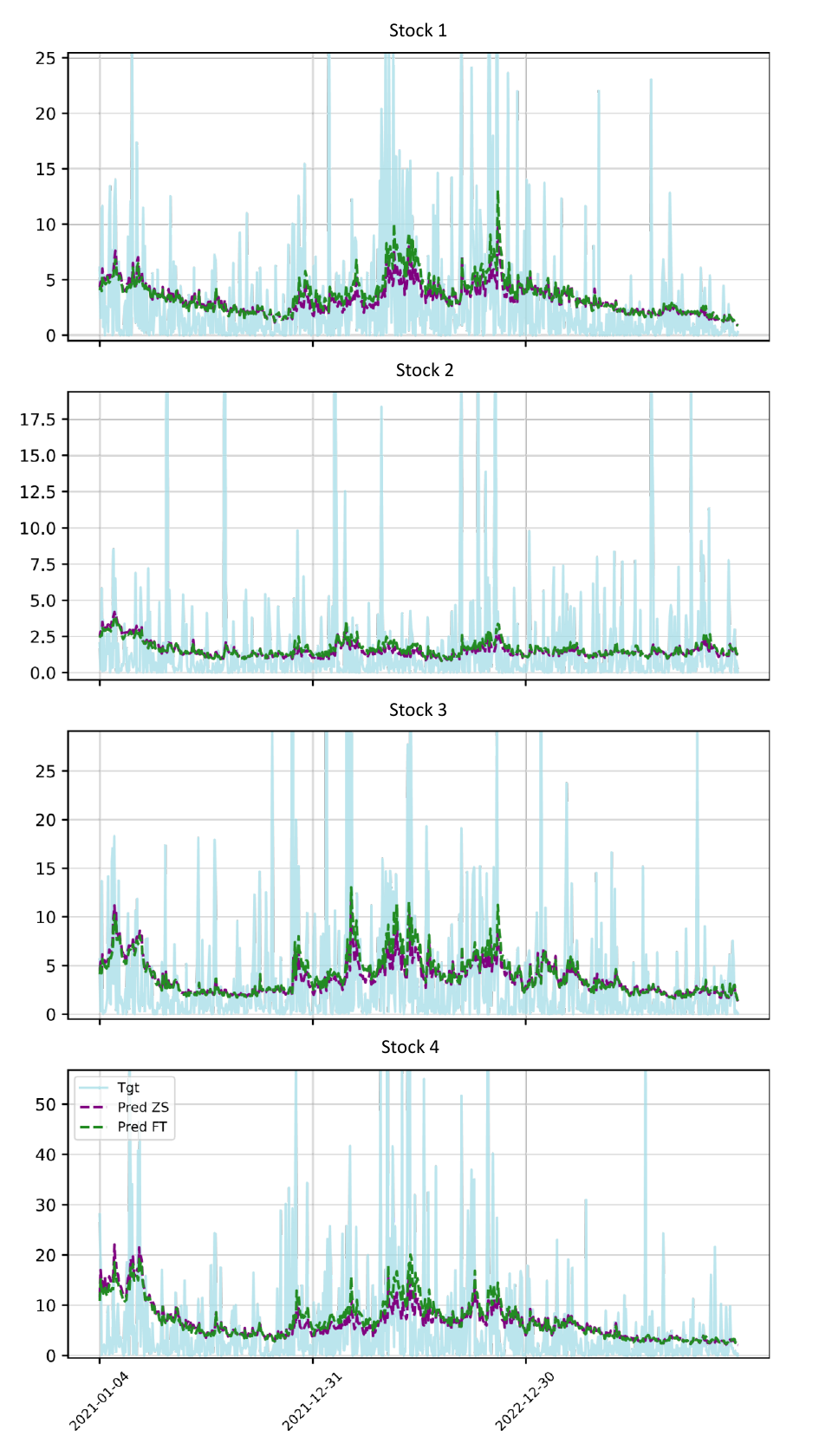}
    \caption{Visualization of fine-tuned forecasts from \methoda on Stock Variance dataset. Note that since sometimes the squared returns are very large, we clip the plot but not the data during training and evaluation.}
    \label{fig:visualization_stocks}
\end{minipage}
\hfill
\begin{minipage}{0.45\textwidth}
    \centering
    \includegraphics[width=\textwidth]{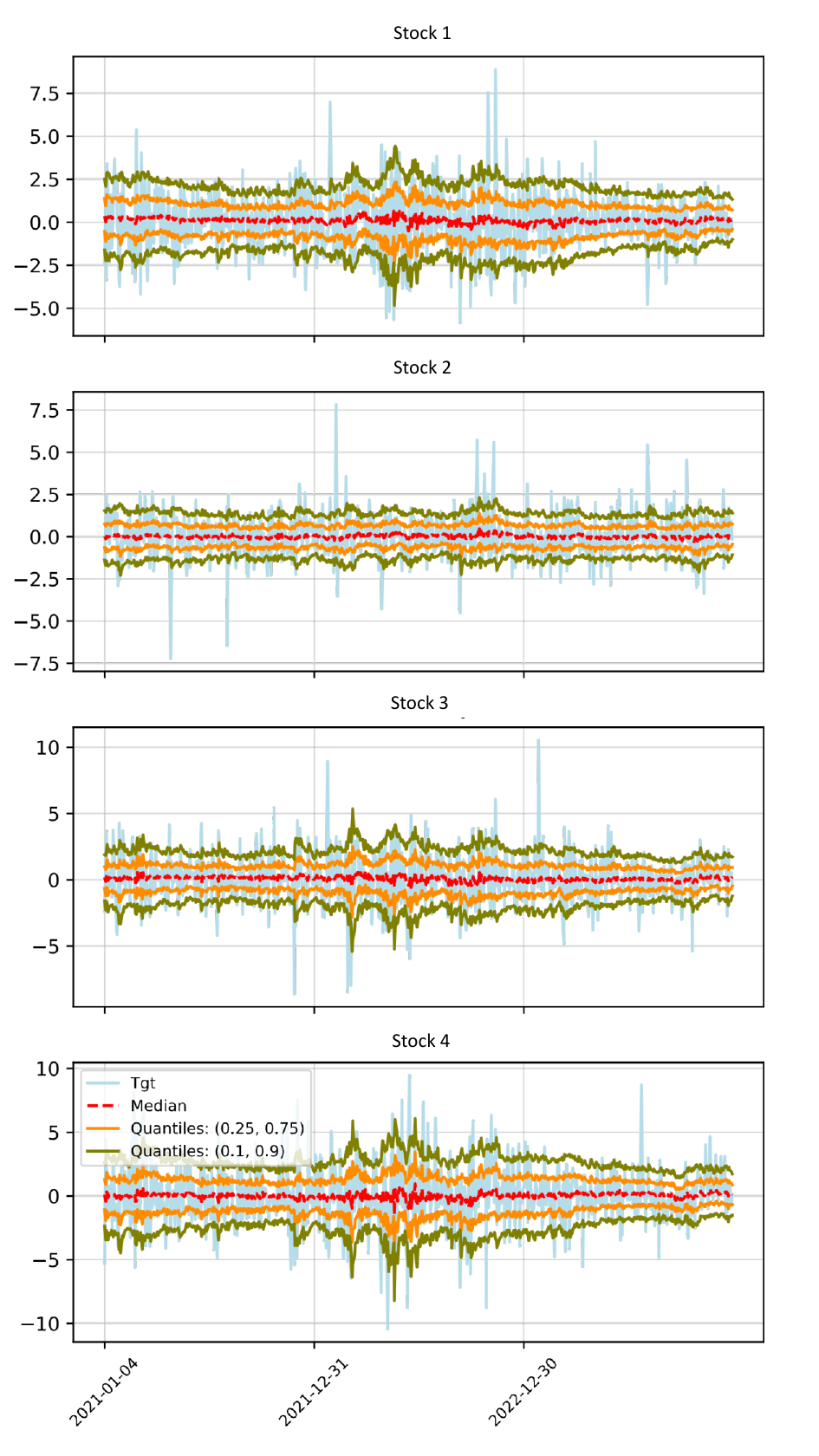}
    \caption{Visualization of fine-tuned probabilistic forecasts from \methoda on Stock NLL dataset.}
    \label{fig:visualization_stocks_nll}
\end{minipage}
\end{figure}

\begin{figure}[ht]
\centering
\begin{minipage}{0.45\textwidth}
    \centering
    \includegraphics[width=\textwidth]{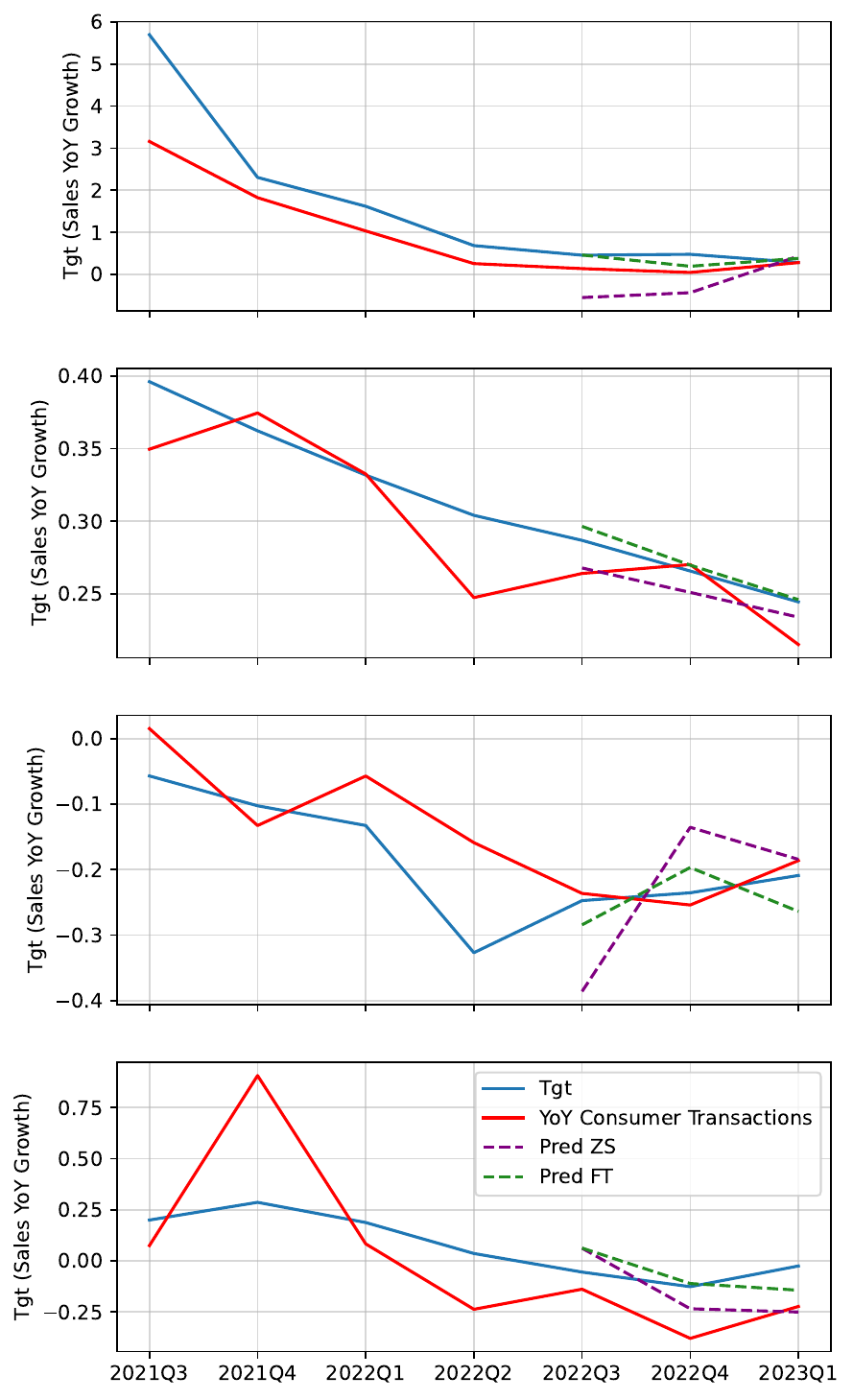}
    \caption{Visualization of fine-tuned forecasts from \methoda on Nowcasting Company Revenue dataset.}
    \label{fig:visualization_nowcasting}
\end{minipage}
\hfill
\begin{minipage}{0.45\textwidth}
    \centering
    \includegraphics[width=\textwidth]{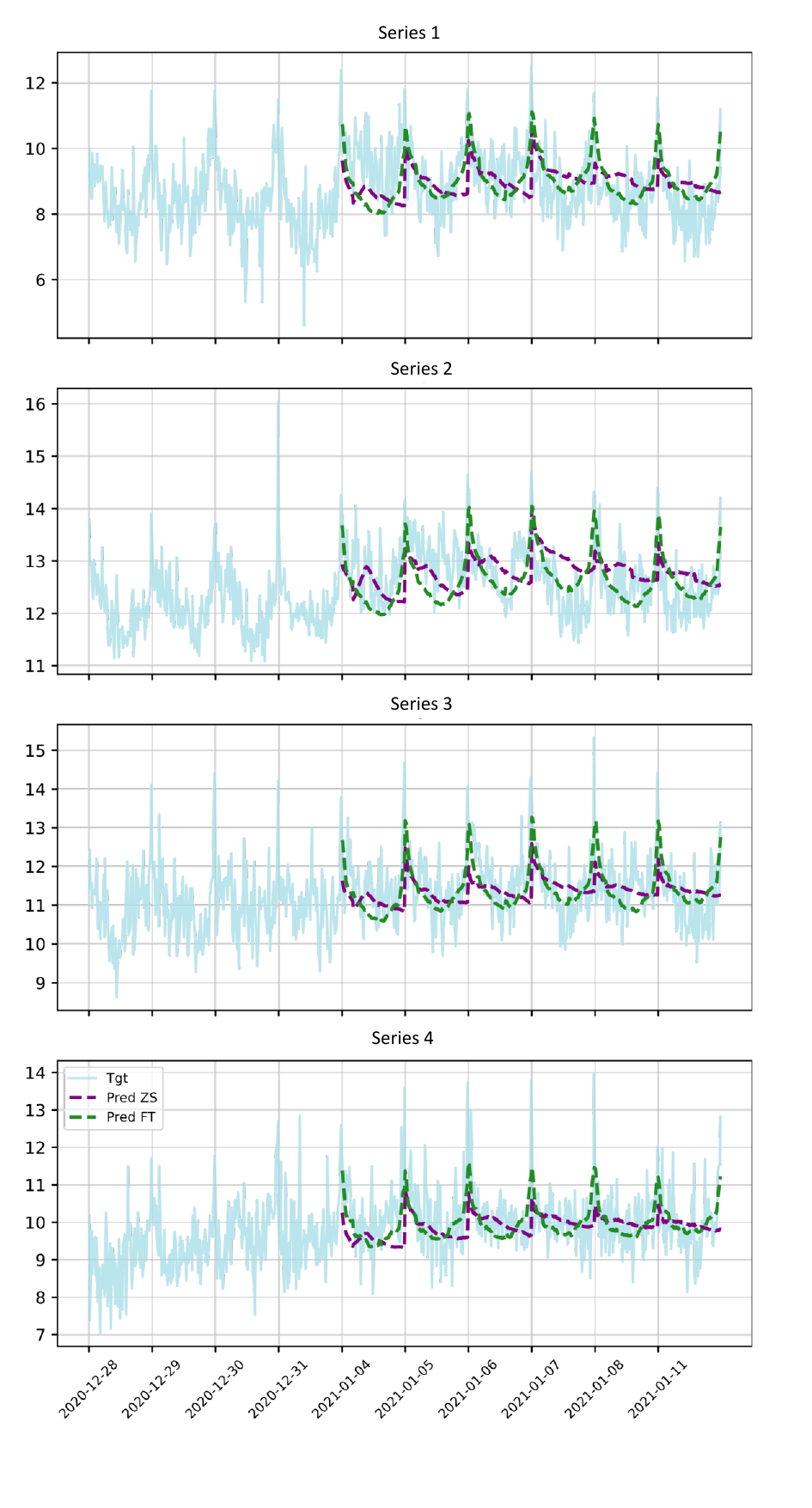}
    \caption{Visualization of fine-tuned forecasts from \methoda on Financial Bars dataset.}
    \label{fig:visualization_bars}
\end{minipage}
\end{figure}

\begin{figure}[ht]
\centering
\begin{minipage}{0.45\textwidth}
    \centering
    \includegraphics[width=\textwidth]{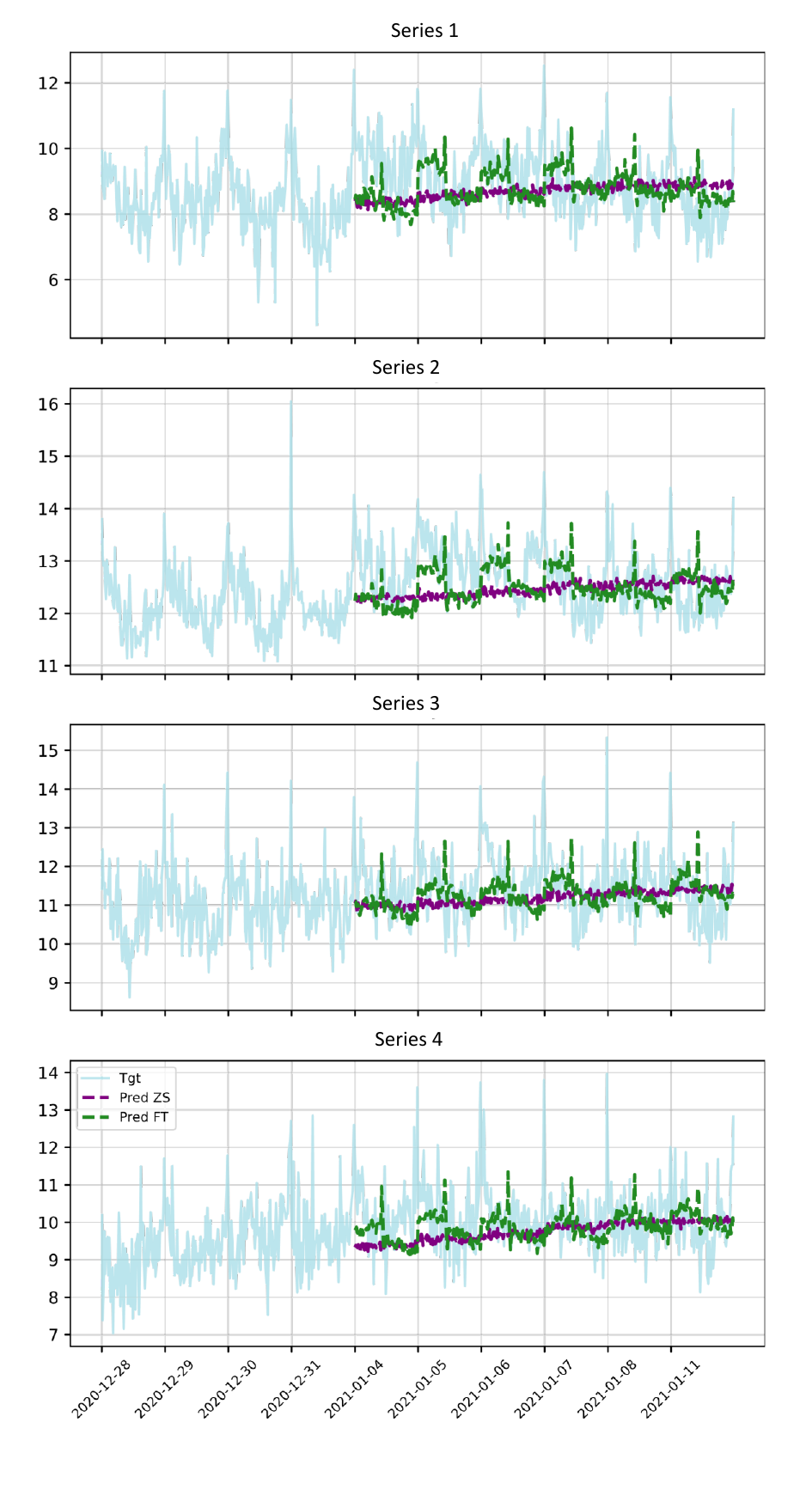}
    \caption{Visualization of fine-tuned forecasts from MOMENT on Financial Bars dataset.}
    \label{fig:visualization_bars_moment}
\end{minipage}
\hfill
\begin{minipage}{0.45\textwidth}
    \centering
    \includegraphics[width=\textwidth]{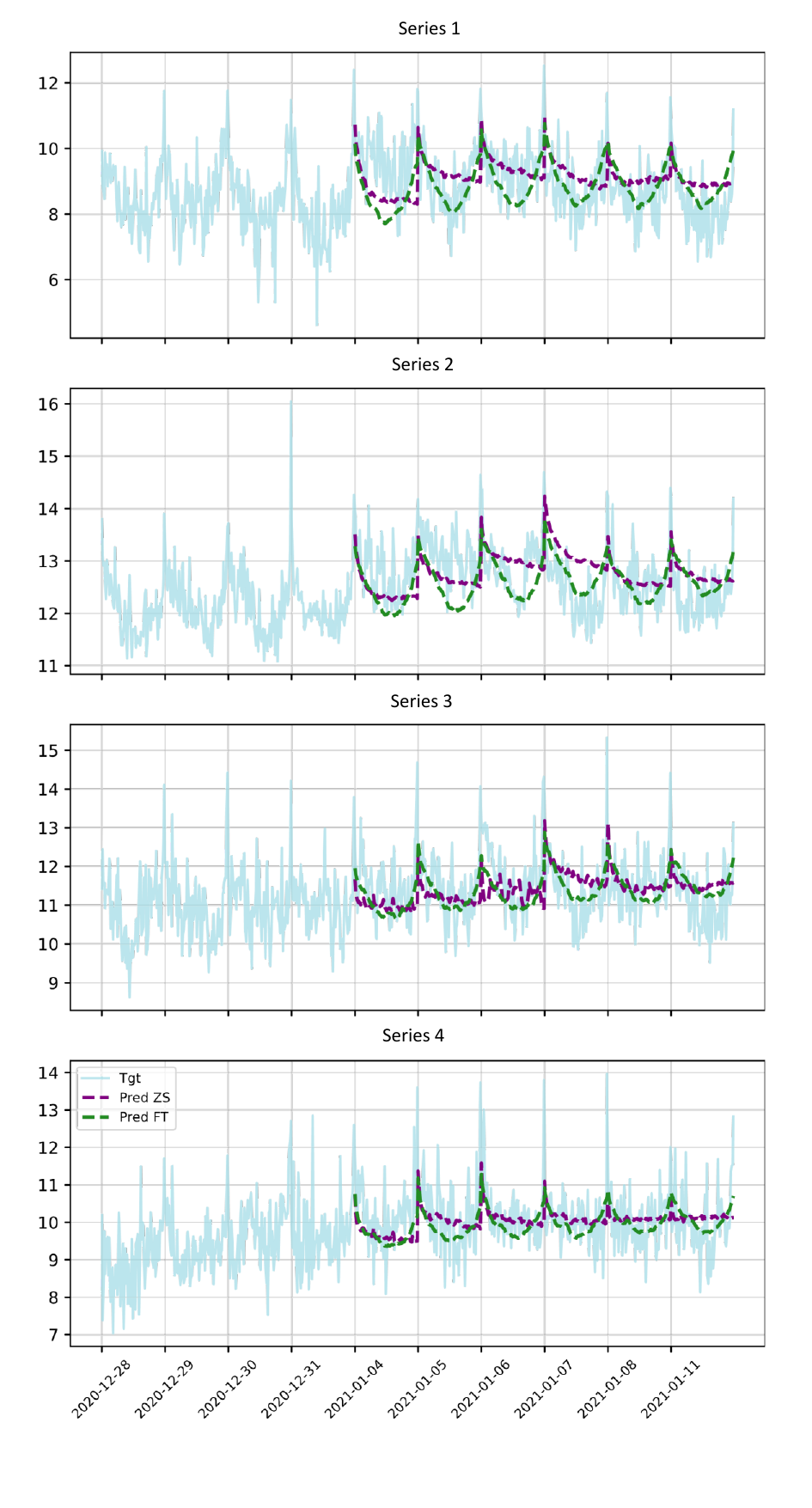}
    \caption{Visualization of fine-tuned forecasts from TTM on Financial Bars dataset.}
    \label{fig:visualization_bars_ttm}
\end{minipage}
\end{figure}

Fig. \ref{fig:visualization_etth1} shows the visualization on ETTh1. Fig. \ref{fig:visualization_stocks} and Fig. \ref{fig:visualization_stocks_nll} show the fine-tuned forecast visualizations on stock variance and NLL. Fig. \ref{fig:visualization_nowcasting} shows the forecast of nowcasting company revenue. Fig. \ref{fig:visualization_bars}, Fig. \ref{fig:visualization_bars_moment} and Fig. \ref{fig:visualization_bars_ttm} show the forecast on bars data using Delphyne, MOMENT and TTM.

\FloatBarrier

\FloatBarrier

\FloatBarrier

\end{document}